\documentclass[fleqn,10pt]{wlscirep}
\usepackage[utf8]{inputenc}
\usepackage[T1]{fontenc}
\usepackage{lineno}
\usepackage{indentfirst}
\usepackage{float}
\usepackage[position=top]{subfig}

\newcommand{\apjl}{%
  Astrophysical Journal, Letters%
}
\newcommand{\apjs}{%
  Astrophysical Journal, Supplement%
}

\newcommand{\aap}{%
  Astronomy and Astrophysics%
}

\title{Constraints on the in-situ and ex-situ stellar masses in nearby galaxies with Artificial Intelligence}

\author[1,2]{Eirini Angeloudi}
\author[1,2]{Jesús Falcón-Barroso}
\author[1,2,3,4]{Marc Huertas-Company}
\author[1,2,5]{Alina Boecker}
\author[1,2]{Regina Sarmiento}
\author[6]{Lukas Eisert}
\author[6]{Annalisa Pillepich}

\affil[1]{Instituto de Astrof\'isica de Canarias, C. V\'ia L\'actea, 1, E-38205 La Laguna, Tenerife, Spain}
\affil[2]{Universidad de la Laguna, dept. Astrof\'isica, E-38206 La Laguna, Tenerife, Spain}
\affil[3]{Universit\'e Paris-Cit\'e, LERMA - Observatoire de Paris, PSL, Paris, France}
\affil[4]{SCIPP, University of California, Santa Cruz, CA 95064, USA}
\affil[5]{
Department of Astrophysics, University of Vienna, T\"urkenschanzstrasse 17, 1180 Vienna, Austria}
\affil[6]{Max Planck Institute for Astronomy,
Königstuhl 17, 69117 Heidelberg, Germany}

\affil[*]{eirini@iac.es}

\begin{abstract}
The hierarchical model of galaxy evolution suggests that the impact of mergers is substantial on the intricate processes that drive stellar assembly within a galaxy. However, accurately measuring the contribution of accretion to a galaxy's total stellar mass and its balance with in-situ star formation poses a persistent challenge, as it is neither directly observable nor easily inferred from observational properties. Here, we present theory-motivated predictions for the fraction of stellar mass originating from mergers in a statistically significant sample of nearby galaxies, using data from MaNGA. Employing a robust machine learning model trained on mock MaNGA analogs (MaNGIA) in turn obtained from a cosmological simulation (TNG50), we unveil that in-situ stellar mass dominates almost across the entire stellar mass spectrum ($10^9M_\odot<M_\star<10^{12}M_\odot$). Only in more massive galaxies ($M_\star>10^{11}M_\odot$) does accreted mass become a substantial contributor, reaching up to 35-40\% of the total stellar mass. Notably, the ex-situ stellar mass in the nearby universe exhibits significant dependence on galaxy characteristics, with higher accreted fractions favored by elliptical, quenched galaxies and slow rotators, as well as galaxies at the center of more massive dark matter halos. %This work sheds new light on the nuanced interplay between galactic morphology, star formation activity, cosmic environment, and the accretion of stellar mass, providing valuable insights into the broader framework of galaxy evolution.
\end{abstract}
\begin{document}

\flushbottom
\maketitle

\thispagestyle{empty}
%\linenumbers

In the currently favored cosmological model, one of the main mechanisms driving the growth of a galaxy's stellar mass and, primarily, the galaxy size is through interactions and mergers with its satellite counterparts\cite{2015ARA&A..53...51S}. This hierarchical process of evolution not only orchestrates the creation of larger structures but also carries crucial consequences in shaping the physical characteristics of galaxies. Mergers are known to trigger structural transformations\cite{2009ApJS..182..216K}, rejuvenate star formation\cite{2011MNRAS.412..591P}, and fuel the activity of Active Galactic Nuclei (AGN)\cite{2008AJ....135.1877E}, as well as contribute to the buildup of massive early-type galaxies (ETGs)\cite{2006ApJ...640..241B}. Their multivariate effect implies that discerning the role of accretion in the assembly of a galaxy's stellar mass  -- particularly in relation to the mass originating from in-situ star formation -- is key for understanding the complex mechanisms that govern galaxy evolution.\looseness-2

However, our understanding of the balance between in-situ and ex-situ mass in a galaxy's total stellar mass remains incomplete.  While the role of internal star formation in the mass assembly of galaxies, peaking around redshift $z \sim 2$, is fairly well established\cite{2014ARA&A..52..415M}, constraining the rate at which galaxy mergers occur across cosmic time and their integrated impact on galaxy evolution is still uncertain. A primary challenge lies in detecting merging events in observational data. Traditionally, this has been accomplished through the identification of pairs of galaxies, sufficiently close both spatially and dynamically, under the assumption that they will merge over a relatively short timescale\cite{2000ApJ...530..660B,2004ApJ...617L...9L}. Alternatively, direct evidence of galaxy interactions, such as tidal tails and structural asymmetries, has been utilized to develop non-parametric methods for the calculation of merger rates across various redshifts\cite{2003AJ....126.1183C,2004AJ....128..163L}. However, both approaches require high quality observational data and carry distinct biases, such as resolution-sensitive thresholds or misidentification of galaxy flybys as merging systems, which may not capture the entire spectrum of merger appearances.

% \cite{2002ApJ...565..208P, 2003MNRAS.346.1189L} for close pairs
%  \cite{2000ApJ...529..886C,} for parametric 

% The recent advances in artificial intelligence have lately reached the field of galaxy evolution, where machine learning techniques have already been successfully employed for a variety of tasks, like morphological classification and galaxy segmentation. But while deep learning can revolutionize merging detection by providing classifications in a fraction of time, its supervised nature, requiring a ground truth acquisition, can act as a impediment. For this reason, deep learning techniques have been used in combination with people-powered research \cite{2018MNRAS.479..415A} or simulation based-inference \cite{2019A&A...626A..49P}. Particularly cosmological simulations have successfully been used to calibrate machine learning models focused on detecting various aspects of mergers, like tidal features \cite{2019MNRAS.483.2968W}, merging phases \cite{2021MNRAS.504..372B}, merger rates \cite{2020ApJ...895..115F} and mergers at higher redshift \cite{2020A&C....3200390C}.
% \cite{2018MNRAS.476.3661D, 2015ApJS..221....8H} for classification
% \cite{2021MNRAS.506..677C} for simulation-based inference

An alternative to direct merger detection is to quantify the integrated effect of mergers in a galaxy, following a more challenging archaeological approach. Since galaxies can only be observed at a single point in cosmic time, tracing their merging history and its impact on a galaxy's evolution is heavily dependent on models or cosmological simulations. Focusing on our cosmic neighborhood, there exist a number of works on deriving the ex-situ stellar mass for a small sample of nearby galaxies through N-body simulations that account for stellar halo buildup only through accretion of dwarf satellite galaxies\cite{2017MNRAS.466.1491H,2022ApJ...930...69S}. Moving further away is an observational challenge, as sufficiently high-quality data is required for unveiling distinctions between the in-situ and the ex-situ stellar populations. So far, parametric surface brightness models have been used in combination with a direct comparison with cosmological simulations\cite{2020MNRAS.491..823B}, revealing that ex-situ stellar mass mainly dominates in the outskirts of galaxies \cite{2010ApJ...709.1018V,2019ApJ...880..111O,2021MNRAS.507.3089D,2023MNRAS.520.5651C}. Particularly for ETGs, several works have attributed the buildup of their outer envelopes to merging events at late times \cite{2013ApJ...768L..28H,2013ApJ...766...47H,2018MNRAS.475.3348H}. However, such results rely on toy models that might not fully capture the more complex patterns tracing the merging history of a galaxy. Machine learning emerges as a promising tool in this regime, having already contributed to the detection of various merger aspects such as tidal features\cite{2019MNRAS.483.2968W}, merging phases\cite{2021MNRAS.504..372B}, and mergers at higher redshift\cite{2020A&C....3200390C}. Deep learning, in particular, holds the potential to revolutionize the understanding of the cumulative impact of mergers on galactic evolution due to its capacity to uncover intricate patterns in complex data. This potential has been showcased in our recent works\cite{2023arXiv231019904E,2023MNRAS.523.5408A}, where neural networks were used to infer key parameters related to the merging history of galaxies in a cosmological simulation context.\looseness-2

% merger rates \cite{2020ApJ...895..115F}
% For that, full spectral fitting has been applied on integral field spectroscopic (IFU) surveys to obtain metallicity and age radial profiles \cite{2004PASP..116..138C, 2017MNRAS.472.4297W}.

% \cite{2014ApJ...795..158M} for radial profiles

Here, we acquire theory-motivated predictions for the integrated impact of merging events in a statistically significant observational sample of nearby galaxies. Building upon our previous work\cite{2023MNRAS.523.5408A}, where we showed that spatially-resolved maps of stellar mass and kinematics can serve as robust predictors of the ex-situ stellar mass fraction of galaxies across two cosmological simulations, we obtain estimations for the amount of stellar mass that has been accreted for $\sim$ 10,000 galaxies in MaNGA. We investigate whether the complex merging histories of galaxies can be traced by secondary dependencies at a fixed stellar mass and attempt to find correlation with the cosmic environment. The novelty of the current work lies in the the fact that it is built upon a robust methodology for the derivation of the ex-situ stellar mass fractions, which allows informed insights on the intricate impacts that mergers can have on galaxy evolution.

\section*{Constraining in-situ and ex-situ stellar mass}

In order to investigate the impact of mergers in the nearby universe across a statistically significant sample, we use the DR17 MaNGA release\cite{2015ApJ...798....7B}, which is the integral field unit (IFU) component of the Sloan Digital Sky Survey (SDSS-IV) (details in Methods). We train a self-supervised machine learning model (see Methods) on MaNGIA\cite{2023A&A...673A..23S}, a sample of mock MaNGA analogs originating from the cosmological simulation TNG50 (details in Methods). MaNGIA not only offers the ground truth that we aim to predict but also includes a desired observational realism specifically matched to MaNGA, rendering it an ideal training sample. Our end goal is to train a neural network that is capable of inferring the global fraction of accreted stars in galaxies from observable IFU maps of stellar mass surface density, stellar velocity and velocity dispersion, as the radial gradients of these properties have already proved to be robust predictors of this quantity in a multi-domain cosmological simulation context\cite{2023MNRAS.523.5408A}. We train a machine learning model on MaNGIA mock galaxies and assess its predictive accuracy on a separate test set reserved from the MaNGIA sample during training. By applying the verified model to the MaNGA galaxies, we are able to acquire predictions for the amount of accreted vs. in-situ stellar mass across the stellar mass range in the nearby universe.\looseness-2

\begin{figure}
    \centering
  \includegraphics[width=0.7\linewidth]{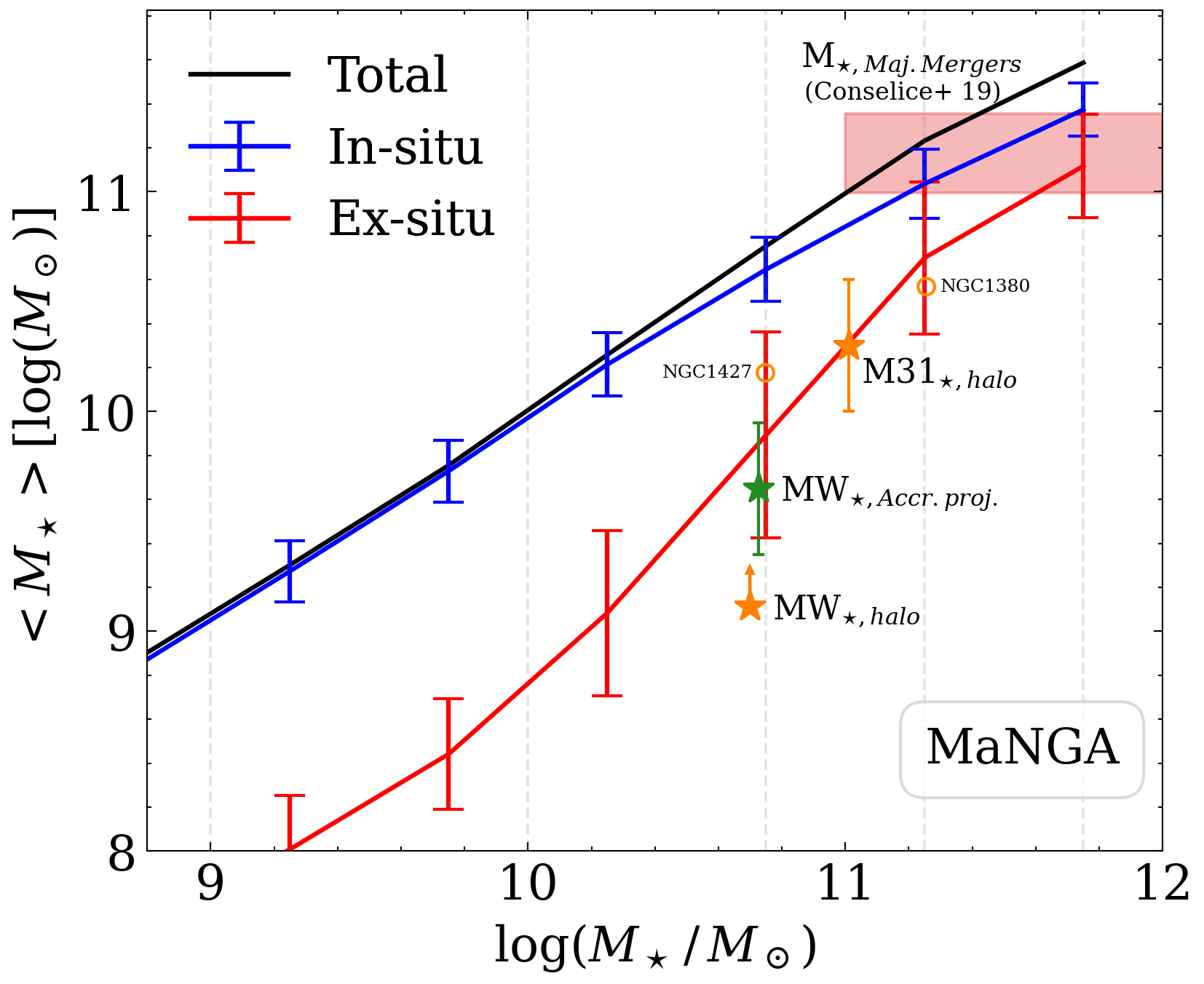}
    \caption{The contribution of the in-situ vs. the ex-situ stellar mass component to the total stellar mass for $\sim$ 10,000 MaNGA galaxies as it is predicted from the Neural Network model trained on the MaNGIA mock dataset. The mean total stellar mass at a given stellar mass bin is drawn with a black solid line and should roughly correspond to a straight line. The average  contribution of the in-situ stellar mass created through internal star formation at each stellar mass bin is drawn with a blue solid line. Correspondingly, the red solid line shows the average contribution originating from accretion/mergers (ex-situ). The error bars contain the 68 per cent of all data. For comparison, we plot the mass of the stellar halo (to distinguish from dark matter halo mass) for Milky Way\cite{2020ARA&A..58..205H} and M31\cite{2017MNRAS.466.1491H} with orange stars (a proxy for accreted mass) and a rough projection of ex-situ stellar mass for Milky Way after it mergers with LMC with a green star (details in discussion). The open circles correspond to an independent ex-situ stellar mass measurement for two early-type galaxies, NGC 1380 and NGC 1427, members of the Fornax galaxy cluster, using population-orbital decomposition methods in combination with cosmological simulation TNG50 from ref.\cite{2022A&A...664A.115Z} The red shaded region corresponds to an independent measurement of mass from major mergers for massive galaxies from ref. \cite{2022ApJ...940..168C} It is apparent that the in-situ stellar mass dominates across the stellar mass range. The ex-situ stellar mass contribution becomes significant only for massive galaxies, revealing a strong dependency of the level of accretion to the stellar mass. Our measurements are roughly compatible with completely independent measurements in the literature.}
    \label{fig:manga_recovery}
\end{figure}

In Fig. \ref{fig:manga_recovery}, we plot the mean in-situ (blue solid line) and ex-situ (red solid line) stellar mass for $\sim$ 10,000 MaNGA galaxies. We find that the in-situ stellar mass dominates across the lower and intermediate stellar mass range, with the ex-situ stellar mass component only becoming significant at the most massive end. More specifically, the contribution of mergers to the stellar mass assembly of low stellar mass galaxies ($M_\star < 10^{10.5} M_\odot$) is nearly negligible, and the entire stellar mass is attributed to in-situ star formation.  As we move to higher stellar masses ($M_\star > 10^{11} M_\odot$), the impact of mergers becomes more relevant, reaching a mean value of the ex-situ stellar mass exceeding $10^{11} M_\odot$ for the most massive galaxies ($M_\star > 10^{11.5} M_\odot$). This indicates that, for such galaxies, the stellar mass originating from mergers constitutes up to 35-40\% of their total stellar mass. Our results are in relative good agreement with previous works in our own galaxy, the local group, and galaxies at the massive end of the stellar mass range, as we will discuss later on.

Overall, we find a pronounced dependency of the amount of ex-situ stellar mass to the stellar mass of the galaxy. Consequently, the more massive a galaxy is the higher the effect of mergers is on its stellar mass assembly. This conforms with the hierarchical build-up scenario in which mergers play an important role in the mass assembly of the most massive galaxies\cite{2010ApJ...725.2312O,2010ApJ...709.1018V}. We emphasize here that this dependency is revealed even though no information about the total stellar mass is provided to the model and any global information is removed by normalizing all input images individually during preprocessing, thus focusing on the radial gradients \cite{2023MNRAS.523.5408A}. The reported correlation additionally exhibits a considerable scatter, suggesting that there exist secondary properties apart from stellar mass that correlate with the integrated merging history of a galaxy, as we aim to investigate in the next section. \looseness-2

\section*{Ex-situ stellar mass dependencies}
We aim to explain the observed scatter in the relationship between accreted stellar mass and total stellar mass by investigating secondary dependencies of the ex-situ stellar mass on galaxy characteristics. Specifically, we examine trends with morphology, star-formation activity, rotation and environment. These properties have already been measured for MaNGA galaxies (see Methods), allowing us to analyze the ex-situ stellar mass dependency within specific subgroups across the stellar mass range. We emphasize here that no prior knowledge of these characteristics was provided as an input to the neural network during training.\looseness-2

\begin{figure*}
    \centering
    \includegraphics[width=0.45\linewidth]{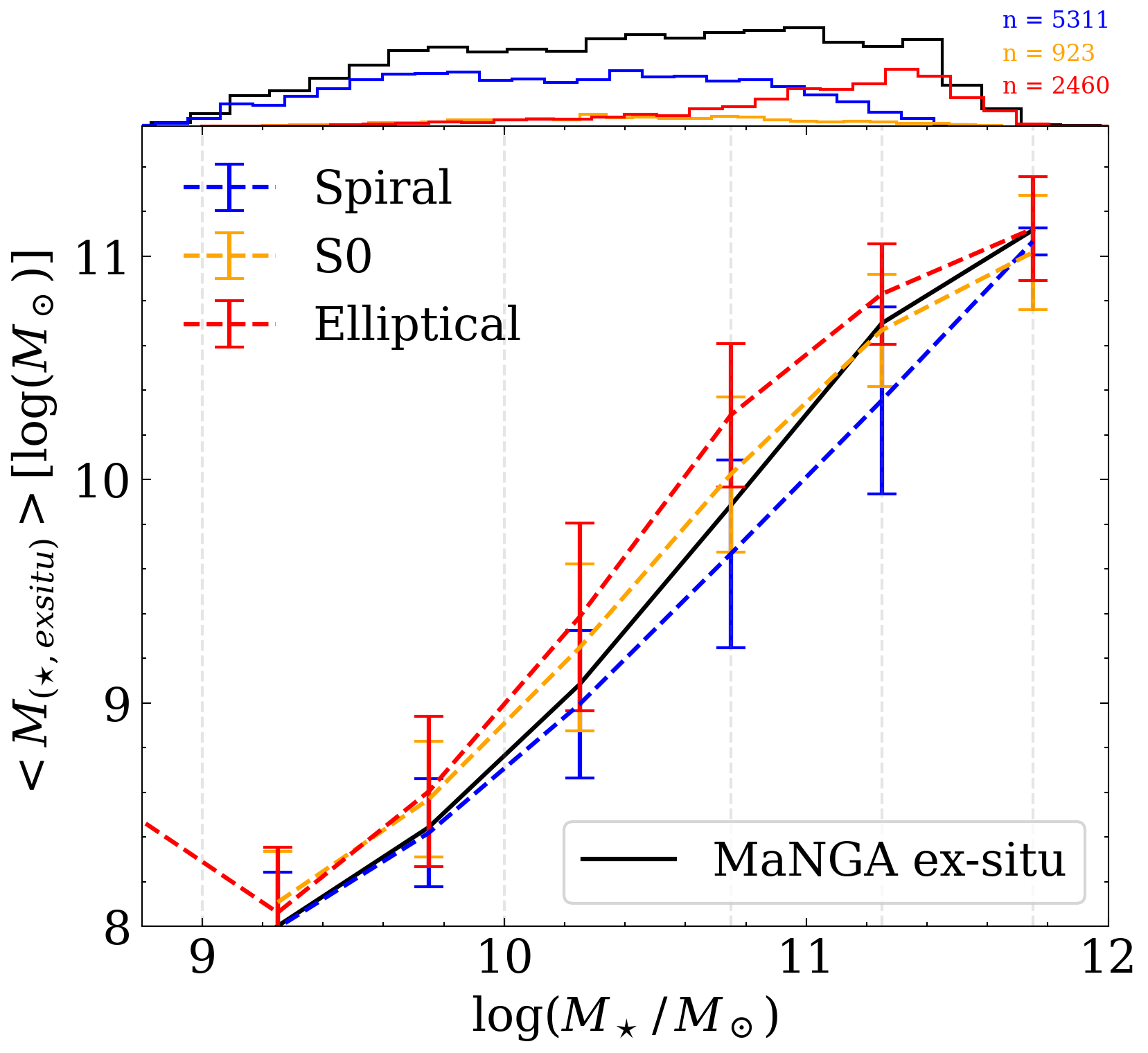}
    \hfil
     \includegraphics[width=0.45\linewidth]{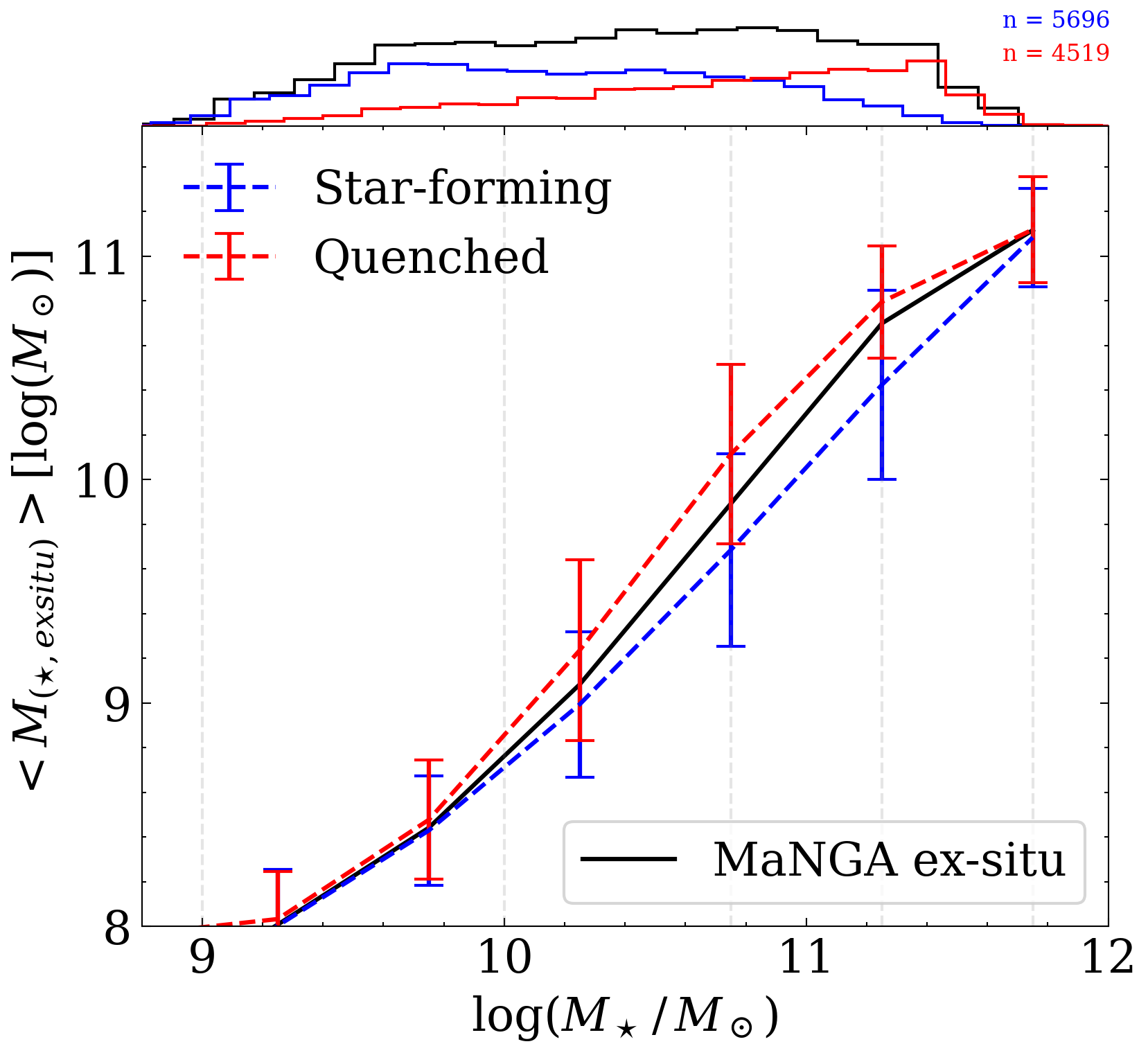}\hfil
\caption{Secondary dependencies of the ex-situ stellar mass with galaxy characteristics. For comparison to the global relation, the mean stellar mass originating from accretion/mergers across the stellar mass range is drawn with a solid black line in both panels. (a) The dependency of the average ex-situ stellar mass on galaxy morphology. At a fixed stellar mass, elliptical galaxies (red dashed line) have more accreted stars than S0 galaxies (yellow dashed line) and significantly more than spiral galaxies (blue dashed line). (b) The dependency of ex-situ stellar mass on star formation. At a fixed stellar mass, quenched galaxies (red dashed line) have accreted a higher fraction of their stellar mass than star-forming galaxies (blue dashed line). The error bars contain the 68 per cent of the data.}
    \label{fig:exsitu_galaxy_dependencies}
\end{figure*}

In Fig. \ref{fig:exsitu_galaxy_dependencies}, the mean ex-situ stellar mass component is plotted for different morphological types and star-formation states. The black solid line represents the mean ex-situ stellar mass fraction of the entire MaNGA sample, while the colored dashed lines correspond to the different subgroups. Our findings reveal that secondary dependencies of the ex-situ stellar mass indeed exist with morphology and star-formation. We find that, on average, elliptical galaxies have accreted more stars than their spiral counterparts at a fixed stellar mass. Interestingly, S0-type galaxies lie between the two other morphological groups, aligning with their expected evolutionary stage and morphological type. A similar pattern emerges for star formation activity, with quenched galaxies exhibiting higher ex-situ stellar mass than star-forming galaxies at a fixed stellar mass. These secondary dependencies are revealed by comparing galaxies at the same stellar mass bins, thus effectively removing the predominant stellar mass dependency. We also note that, while there exists a high overlap in the observed scatter between the different subgroups, the mean values of ex-situ stellar mass consistently separate the distinct classes almost across the entire stellar mass range.\looseness-2

 % This result highlights that the neural network model identified features corresponding to the merging history of a galaxy that correlate with its morphology and star-formation state, even when such characteristics were not explicitly included in the training inputs. 

% \subsection*{Galaxy rotation}

\begin{figure*}
    \centering
     \includegraphics[width=0.7\linewidth]{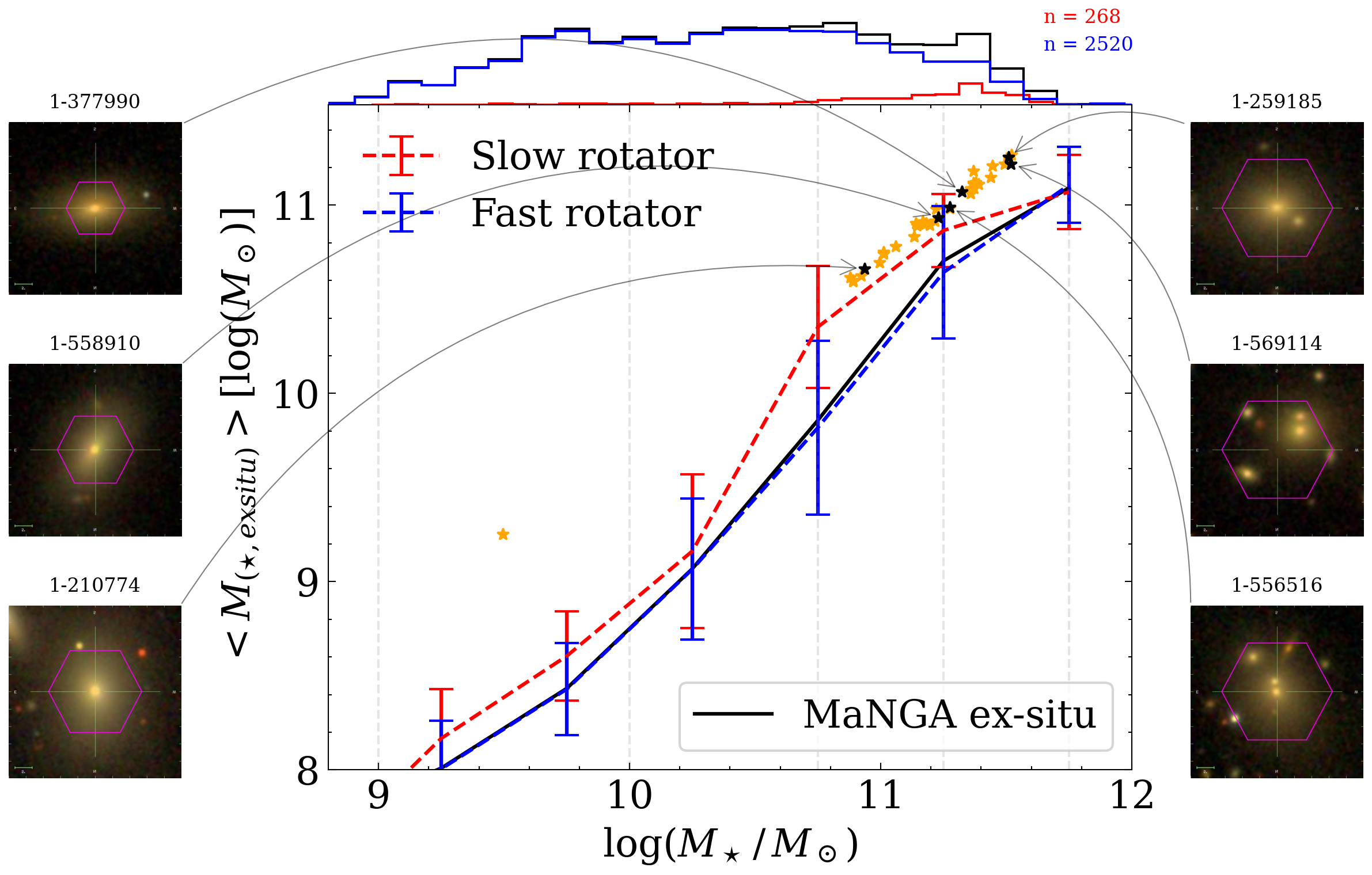}
     \hfil
    \caption{Dependencies of the ex-situ stellar mass on rotation. For comparison to the global relation, the mean stellar mass originating from accretion/mergers across the stellar mass range is drawn with a solid black line. At a fixed stellar mass, slow rotators (red dashed line) exhibit higher ex-situ stellar mass than their fast rotator counterparts (blue dashed line). The error bars contain the 68 per cent of the data. The star points correspond to outliers, galaxies classified as fast rotators with a predicted fraction of ex-situ stellar mass > 0.5. We plot 6 SDSS images of the outliers' sample and identify several intriguing actual fast rotators (left-hand side images) as well as galaxies undergoing a merger event (right-hand side images). The image labels correspond to the MaNGA ID of each galaxy.}
    \label{fig:exsitu_rotation}
\end{figure*}

We also find that slow rotators\cite{2016ARA&A..54..597C} have accreted a higher fraction of their stellar mass, on average, than fast rotators (Fig. \ref{fig:exsitu_rotation}). While our statistics for slow rotators are not extensive, especially in lower masses, their mean ex-situ stellar mass consistently surpasses the global relation across all stellar mass bins. Given the limited size of our slow rotators sample, we also opt to investigate peculiar objects - galaxies classified as fast rotators but exhibiting an unexpectedly high ex-situ stellar mass fraction (> 0.5). Such objects could be anomalies due to erroneous processing or genuinely intriguing objects managing to maintain their fast rotation even after significant merging events. We identify  $\sim40$ galaxies with these characteristics (highlighted as orange and black stars in Fig. \ref{fig:exsitu_rotation}). Upon closer examination, we find that a significant proportion of them appear to be indeed fast rotators, compact galaxies with a substantial bulge (left-hand side of the SDSS images in Fig. \ref{fig:exsitu_rotation}). To further examine these interesting objects, deeper observations might be required. Additionally, we also identify on-going merging events in the outliers data (right-hand side of the SDSS images in Fig. \ref{fig:exsitu_rotation}), pointing to a erroneous calculation of their rotational properties. Notably, the specific objects have also been flagged as unclean in their fast-rotator classification, indicating an additional visual categorization as merging events.

\begin{figure*}
    \centering
    \includegraphics[width=0.45\linewidth]{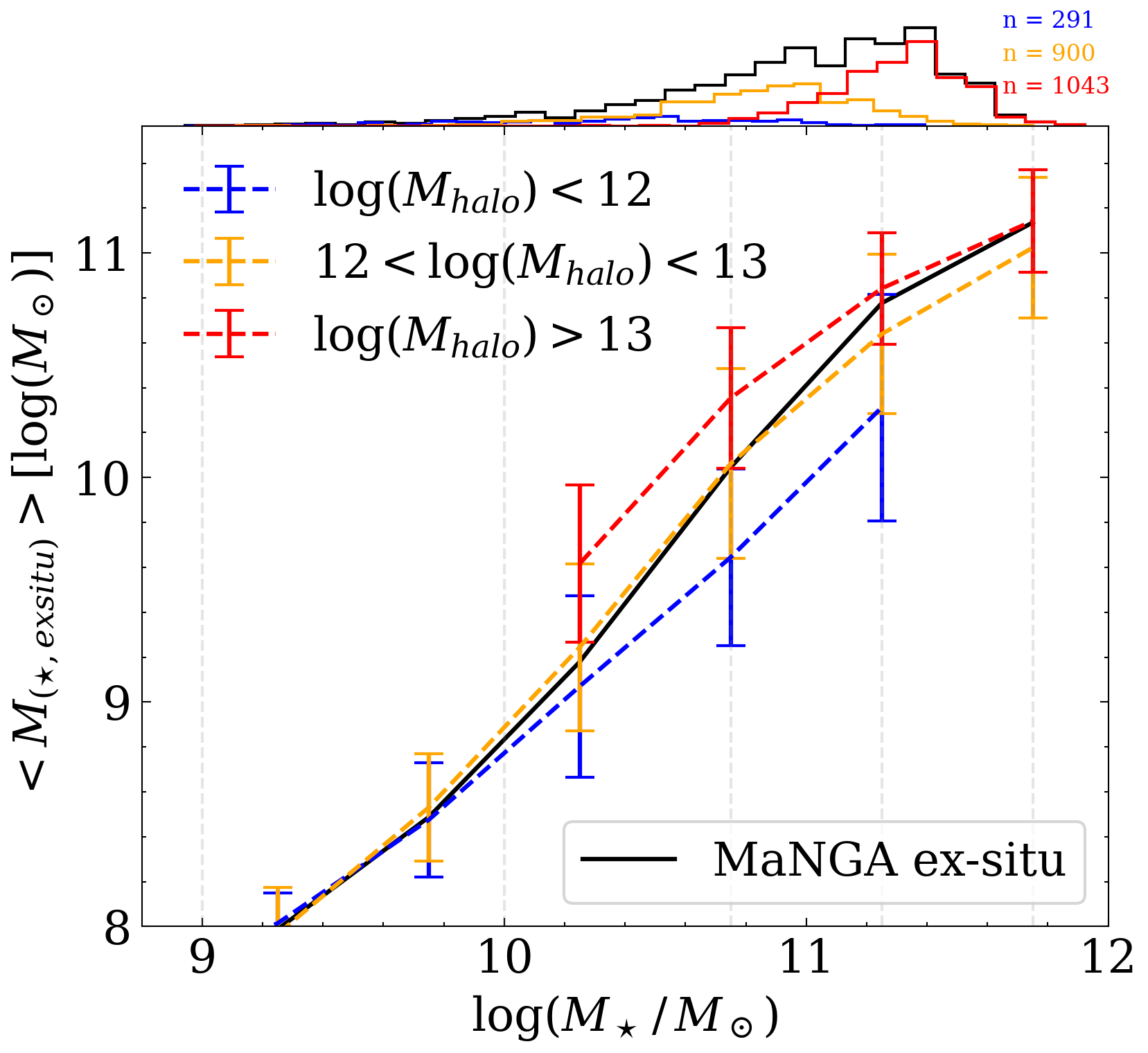}
    \hfil
    \includegraphics[width=0.45\linewidth]{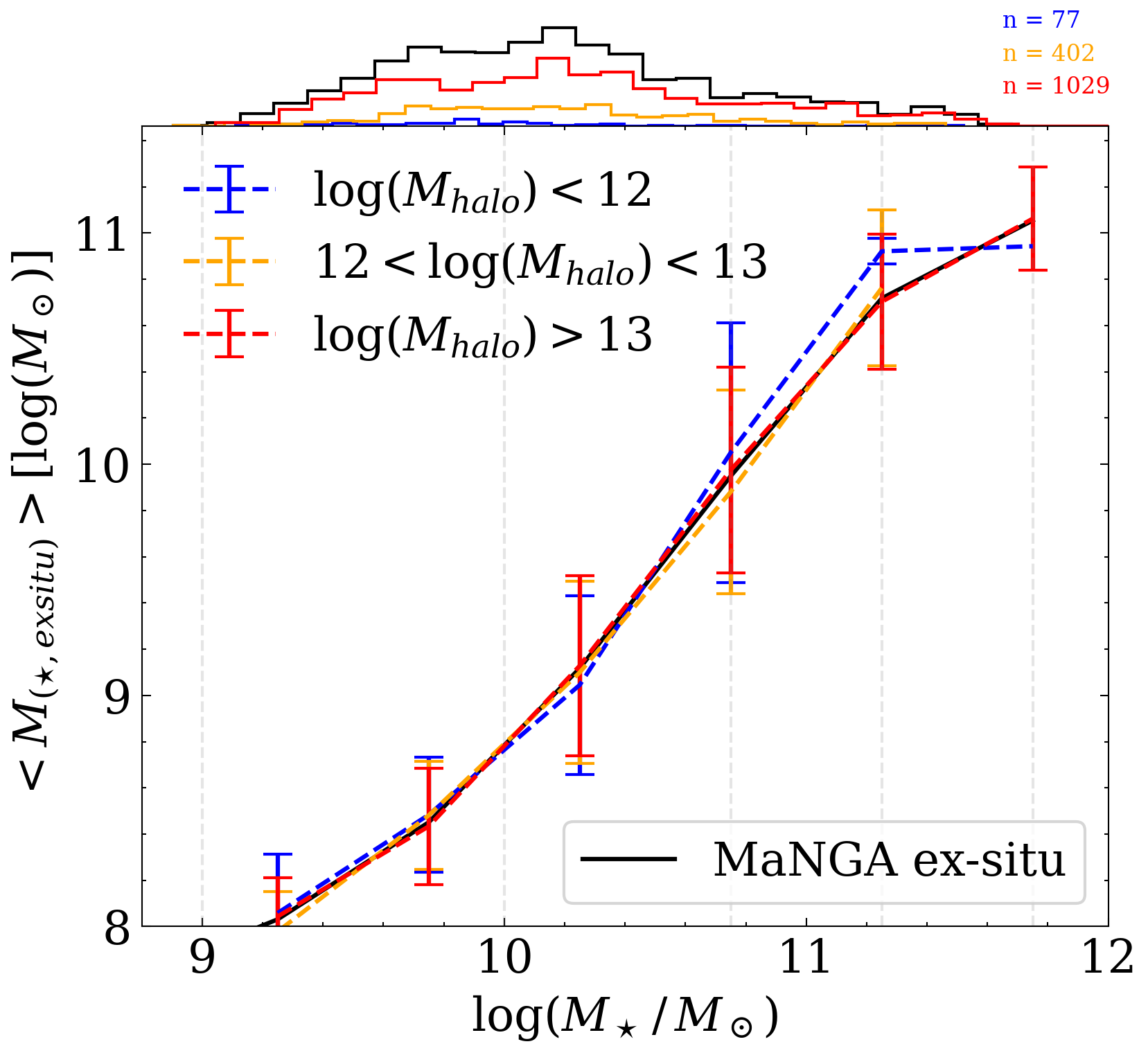}

\caption{Dependencies of the ex-situ stellar mass on environment. For comparison to the global relation, the mean stellar mass originating from accretion/mergers across the stellar mass range is drawn with a solid black line in both panels for central and satellite galaxies respectively. (a) The dependency of ex-situ stellar mass on the dark matter halo mass for central galaxies. At a fixed stellar mass, galaxies at the centre of more massive haloes (red dashed line) have accreted higher fraction of their stellar mass. (b) The dependency of ex-situ stellar mass on the dark matter halo mass for satellite galaxies. The fraction of ex-situ stellar mass in satellite galaxies is not impacted by the dark matter halo mass at a fixed stellar mass. The error bars contain the 68 per cent of the data.}
    \label{fig:exsitu_env_dependency}
\end{figure*}

% \subsection*{Effect of host halo properties}
Finally, we aim to investigate the impact of the environment on the merging history of galaxies (Fig. \ref{fig:exsitu_env_dependency}). Previous studies have reported correlations between the stellar population properties and the mass of the host dark matter halo that a galaxy resides in\cite{2015ApJ...807...11G,2022ApJ...933...88O,2024NatAs.tmp...42S}. This prompts us to explore whether such correlations extend to the merging history of galaxies. Using the self-calibrating halo-based galaxy group finder on SDSS from ref\cite{2021ApJ...923..154T}, we find that galaxies in the center of more massive dark matter haloes have accreted more stellar mass than their counterparts with a similar total stellar mass but located in the center of less massive dark matter haloes (isolated galaxies have been removed from this analysis for more reliable halo mass estimates). This result signifies that the integrated merging history of a galaxy is not only correlated with its stellar mass but also with the properties of its host dark matter halo. Interestingly, we find that this dependency is not present for satellite galaxies (right panel of Fig. \ref{fig:exsitu_env_dependency}).

%Correspondingly, we also find that central galaxies with a higher number of satellite galaxies exhibit higher ex-situ stellar mass components than central galaxies with fewer satellites at a fixed stellar mass.

% We emphasize here that no prior knowledge on halo mass or central vs. satellite classification was included during the training of the model. \looseness-2 

% This result signifies that the integrated merging history of a galaxy is not only correlated with its stellar mass but is also influenced by the properties of the host dark matter halo.
%Similar to the presentation in Fig. \ref{fig:exsitu_galaxy_dependencies} and Fig. \ref{fig:exsitu_rotation}, the subdivisions of the mean ex-situ stellar mass in MaNGA corresponding to different host halo masses and satellite abundances is plotted in Fig. \ref{fig:exsitu_env_dependency}. \looseness-2 

% \begin{figure}
%     \centering
%     \includegraphics[width=0.45\linewidth]{Figures/Outliers/6.png}\hfil
%     \includegraphics[width=0.45\linewidth]{Figures/Outliers/27.png}\\
%     \includegraphics[width=0.45\linewidth]{Figures/Outliers/0.png}\hfil
%     \includegraphics[width=0.45\linewidth]{Figures/Outliers/23.png}\\
%     \includegraphics[width=0.45\linewidth]{Figures/Outliers/15.png}\hfil
%     \includegraphics[width=0.45\linewidth]{Figures/Outliers/12.png}
%     \caption{Outliers fast rotators with high exsitu fractions.}
%     \label{fig:outliers}
% \end{figure}

\section*{Discussion}

We calculate physics-based measurements on the integrated impact of mergers in the nearby universe for a statistically significant sample of galaxies. We achieve so by taking advantage of the power of machine learning to determine the relative contributions of the ex-situ and in-situ stellar mass components to the total stellar mass of galaxies using IFU MaNGA data. Our findings reveal that galaxies with low stellar mass ($M_\star < 10^{10.5} M_\odot$) are mainly comprised of in-situ stars with an almost negligible ex-situ component. On the other hand, we find that the impact of mergers is more prominent for galaxies at higher stellar masses ($M_\star > 10^{11} M_\odot$), where the ex-situ stellar mass component can reach up to 35-40\% of the total stellar mass.

%Our findings reveal that in-situ star formation predominantly drives stellar mass growth in galaxies with lower and intermediate stellar masses, with the contribution of mergers to the stellar component of galaxies only becoming significant at the high end of the stellar mass range. More specifically, w

Our findings allow us to put constrains on the integrated amount of stellar mass originating from in-situ star formation in comparison to the component accreted from mergers and interactions in the local universe. By integrating the contribution of in-situ and ex-situ stellar mass across our MaNGA sample, we find that $\sim 70\%$ of the total stellar mass in the nearby universe is attributed to in-situ star formation by our model and only $\sim 30\%$ is attributed to accretion through mergers and interactions. In a volume-limited sample, attained with volume corrections as reported in ref\cite{2017AJ....154...86W}, the fraction of ex-situ mass drops even further to $\sim 19\%$. This result greatly underlines that, while the contribution of mergers to the mass assembly of galaxies is important, it remains a secondary factor in stellar mass growth in comparison to in-situ star formation. Notably, our findings fall partially short of predictions from simulations, such as the Illustris cosmological simulation estimating a global ex-situ stellar mass fraction of 30\% at redshift z = 0 \cite{2016MNRAS.458.2371R}, and the semi-analytical model GALFORM predicting a corresponding value of 40\%\cite{2023MNRAS.518.5323H}.

The constraints set in the current work are in relatively good agreement with estimations performed for our own galaxy and the local neighborhood. Particularly for Milky Way, the component primarily attributed to the accretion of stars from satellite galaxies is identified as the Milky Way halo \cite{2008ApJ...680..295B,2020ARA&A..58..205H}, estimated to have a stellar mass of approximately $\sim 1.3 \times 10^{9}$ \cite{2020ARA&A..58..205H}. This serves as a lower limit for the amount of accreted mass for the Milky Way, falling below the mean relation of ex-situ stellar mass recovered for MaNGA galaxies (refer to the orange star in Fig. \ref{fig:manga_recovery}). For a speculative future scenario, we introduce an additional point (green star) representing the rough projected ex-situ stellar mass estimate when the Milky Way eventually merges with its most massive satellite, the Large Magellanic Cloud (LMC), with a calculated stellar mass of $\sim 1.5 \times 10^{9}$\cite{2012AJ....144....4M}. Interestingly, this projection aligns more closely with our mean ex-situ stellar mass relation. However, it's important to note that the Milky Way is believed to have had a rather atypical assembly history given its mass\cite{2019MNRAS.486.3180K}. We also plot the estimated halo mass of M31\cite{2017MNRAS.466.1491H} and find that it lies on our mean ex-situ stellar mass relation, being consistent with the constraints found in the current work.

Furthermore, an additional validation of our results at the high end of the stellar mass range can be obtained through a comparison with findings from studies measuring the rates at which galaxies merge at various redshifts\cite{2022ApJ...940..168C,2017MNRAS.470.3507M}. Such works offer a completely independent observational measurement of the ex-situ stellar mass, which is based on the calculation of the fraction of galaxies found in close pairs in extended survey samples. By tracking the number of merging events per unit co-moving volume for redshift $z < 3$, one can integrate the total accreted stellar mass in this redshift range when the average stellar mass of the infalling galaxy is known. Specifically in ref.\cite{2022ApJ...940..168C}, the authors approximate that for massive galaxies ($M_\star > 10^{11} M_\odot$) the amount of ex-situ stellar mass attributed to major merger events for redshift $z < 3$ is $1.48^{+ 0.78}_{-0.49} \times 10^{11} M_\odot$ (plotted in Fig. \ref{fig:manga_recovery} as a shaded red region). This is in relative good agreement with our results in the high end of the stellar mass range, taking all the uncertainties involved in the calculations under consideration. Additionally, an independent ex-situ stellar mass measurement for two external early-type galaxies, NGC 1380 and NGC 1427, members of the Fornax galaxy cluster, has been reported using population-orbital decomposition methods in combination with the TNG50 cosmological simulation from ref.\cite{2022A&A...664A.115Z}. These measurements also align closely with our relation (plotted in Fig. \ref{fig:manga_recovery} as open circles).

Diving further, we find a very prominent dependency of the amount of accreted stellar mass to the total stellar mass of the galaxy. This generally agrees with results from cosmological simulations\cite{2012MNRAS.425..641L,2016MNRAS.458.2371R,2020MNRAS.497...81D,2019MNRAS.487.5416T}, zoom-in simulations\cite{2010ApJ...725.2312O} and semi-analytical models\cite{2013ApJ...766...38L}. Across these studies, there is a consensus that the more massive a galaxy is, the higher the fraction of its stellar mass originating from accretion and mergers.
Similar to this work, simulations predict almost minimal ex-situ stellar mass for galaxies with stellar mass $M_\star < 10^{10.5}$ and  a steep increase of the accreted stellar mass fraction for stellar masses above $10^{10.5}$.
However, apart from ref.\cite{2012MNRAS.425..641L}, all these models predict very high ex-situ stellar mass fractions of the orders of $\sim 80\%$ for the most massive galaxies ($M_\star > 10^{11.5} M_\odot$), levels that are not reached in our results.  This discrepancy may be attributed to the fact that our MaNGIA training sample originates from the limited-volume TNG50 cosmological simulation. While TNG50 provides finer resolution, it is confined to a smaller simulation box, resulting in massive galaxies lying on average in lower-mass haloes compared to simulations of a larger volume. However, since our MaNGIA sample is matched to MaNGA, with a selection criterion that also does not favor central galaxies in very massive haloes\cite{2017AJ....154...86W}, we consider this limitation acceptable and focus primarily on the statistical interpretation of the results.

Apart from the strong dependency of ex-situ stellar mass to the total stellar mass of a galaxy, we additionally observe that the said relation exhibits a significant scatter. Attempting to explain it, we find that the amount of stellar mass originating from mergers in a galaxy has secondary dependencies on morphology, star-formation, rotation, and environment. More specifically, we find that elliptical galaxies have accreted a higher fraction of their stellar mass than spiral galaxies at a fixed stellar mass. This is not a surprising result, as merging is known to be a possible mechanism for the creation of elliptical galaxies\cite{2006MNRAS.366..499D,2009ApJS..182..216K,2010ApJ...709.1018V} and similar trends are also produced in cosmological simulations\cite{2017MNRAS.467.3083R,2019MNRAS.487.5416T}. Additionally, we find that galaxies with quenched star formation have accreted a greater fraction of their stellar mass, on average, than their star-forming counterparts. This result aligns with previous studies that attribute fast quenching to gas-rich mergers, which can induce intense bursts of star formation and trigger AGN activity, depleting galaxies from their gas content\cite{2018MNRAS.473.1168R,2019ApJ...874...17B}. %Additionally, since the ex-situ stellar mass component measured in our study is also a good tracer of the dry mergers' effect on the total stellar mass growth. Quenched galaxies are thought to increase their stellar mass and size mainly due to dry mergers \cite{2009ApJ...699L.178N, 2015ApJ...813...23V}, further underling the pivotal role of merging in the evolution of massive elliptical systems. 

% This finding offers insights into the intricate processes driving the quenching of star formation in galaxies, including gas depletion \cite{1998ARA&A..36..189K} and the influence of AGN and stellar feedback \cite{2010MNRAS.402.1536S}.
 
%at fixed stellar mass, higher values of S/T and C82 correlate with higher-than-average ex situ stellar mass fractions.

We identify an additional secondary dependency of the ex-situ stellar mass on galaxy rotation. Despite the limited sample of slow rotators in this study, our analysis reveals that slow rotators exhibit a higher ex-situ stellar mass component than fast rotators at a fixed stellar mass. This result agrees with previous studies that attribute disturbed rotational properties to merger events\cite{2014MNRAS.444.3357N}. Imprints of mergers and interactions, such as tidal features, have been shown to prevail in slow rotators in comparison to fast rotators, particularly in ETGs\cite{2022ApJ...925..168Y}.  Moreover, our approach can serve as a valuable tool for detecting anomalies in the data, enabling the identification of rare objects, such as fast-rotating galaxies with unusually high ex-situ stellar mass components.\looseness-2

%Our results underline the effect of mergers on the kinematic properties of galaxies for the first time in a statistically significant observational sample.

Finally, we attempt to assess the impact of the environment and in particular the host dark matter properties on the merging history of a galaxy to get further insights into the interplay between the two. Interestingly, we find that the contribution of ex-situ stellar mass to the total mass is higher for central galaxies located at the center of more massive dark matter halos than their counterparts in less massive halos. Intriguingly, this trend is not mirrored in satellite galaxies, suggesting that their merging history is not equally influenced by their host halo properties. Previous works have found relations between environment and merger rates, especially dry mergers in over-density regions\cite{2010ApJ...718.1158L}, but the dependence of merger rates on the halo mass was not conclusive\cite{2012ApJ...754...26J}. Notably, a recent study\cite{2022ApJ...930...69S} identified a tight relation between the number of satellites and the accreted mass for several galaxies in our local neighborhood, further supporting our results.

%Additionally, we find a similar trend for the ex-situ stellar mass with the number of satellites surrounding a central subhalo, where a greater number of satellites results in a higher fraction of accreted stellar mass. 
% that a central galaxy is more prone to mergers when located in a halo with a higher count of satellite galaxies.

All in all, this work provides theory-motivated observational constraints on the balance of in-situ vs. ex-situ stellar mass in the mass assembly of galaxies in the nearby universe for a statistically significant sample. We find that the impact of mergers on a galaxy's stellar assembly, while overall secondary in comparison to in-situ star formation, is more prominent at the high mass end of galaxies and less significant at the low and intermediate stellar masses. Our findings additionally reveal secondary dependencies of the fraction of accreted stellar mass with galaxy morphology, star-forming activity, rotational properties, as well the galaxy's environment throughout the entire stellar mass range. Looking ahead, the upcoming high-redshift surveys, combined with the computational capabilities of machine learning, hold the potential to unravel the balance between in-situ and ex-situ stellar mass evolution across cosmic time.

%Moreover, we observe a noteworthy influence of a galaxy's environment on its merging history, particularly for central galaxies situated at the core of their host dark matter halos.

\section*{Methods}

\subsection*{Overall approach}

The goal of the present work is to acquire estimations for the contribution of in-situ vs. ex-situ stellar mass for an observational sample of the local universe that is statistically meaningful. Since the effect of mergers (ex-situ stellar mass) is neither directly observable nor straightforward to obtain from a galaxy's observational properties, we need to resort to cosmological simulations for a view on the merging histories of nearby galaxies. We choose to train a deep learning model to predict the fraction of ex-situ stellar mass from observable spatially-resolved maps in simulated galaxies. The first challenge that might arise from such an approach is that training on a particular cosmological simulation can induce bias in the predictions of the model, which can in turn be carried on the application of the model on real galaxies. Moreover, observational data contain inherent observational effects and noise absent in idealized simulated data, complicating the direct application of a model trained on raw simulation data to observational images.

To address these issues, our overall approach is built as follows: we use two state-of-the-art cosmological simulations instead of one to cross train our neural network model, so that we can identify a set of inputs that is a robust predictor of the ex-situ stellar mass fraction across the distinct sub-grid physics that govern each simulation. In our previous work \cite{2023MNRAS.523.5408A}, we have shown that the gradients of the spatially-resolved maps of stellar mass, stellar velocity and stellar velocity dispersion are accurate and robust predictors of the global ex-situ stellar mass fraction of nearby galaxies across the TNG100 and the EAGLE cosmological simulations. However, the trained models are still missing the required realism for application on real galaxies. Since our end goal is to investigate the impact of mergers on the mass assembly of galaxies in the local universe in a statistical manner, we decide to employ the machine learning model on MaNGA \cite{2015ApJ...798....7B}, the currently largest integral field spectroscopic survey containing data for $\sim$ 10,000 nearby galaxies. To train the aforementioned model, we require a mock dataset derived from simulations with a realism matched to MaNGA, which is also statistically significant. We  turn to MaNGIA (Mapping Nearby Galaxies with IllustrisTNG Astrophysics), the largest mock dataset to date designed to mimic the MaNGA survey. By training and validating the predictive capabilities of the machine learning model on MaNGIA, we can confidently apply it to MaNGA galaxies and obtain a first prediction on the ex-situ vs. in-situ stellar mass contribution to their total stellar mass.

The machine learning model takes as inputs the robust IFU maps of stellar mass density and kinematics and produces as an output a prediction of the ex-situ stellar mass fraction for each galaxy. This selection of inputs is derived from the cross-training across TNG100 and EAGLE in the initial step of our methodology. Since MaNGIA originates from the TNG50 cosmological simulation, we expect these inputs to generalize well when applied to both mock and real data. The ground-truth of the ex-situ stellar mass is available for MaNGIA, as the stellar assembly catalogs have been resolved for the TNG50 simulation \cite{2017MNRAS.467.3083R}. These catalogs provide, among others, a measurement of the ex-situ stellar mass for each galaxy, by traversing its merger tree and classifying all stars formed in other galaxies other than the main progenitor branch (that either merged or interacted with the main galaxy) as accreted.  Regarding the inputs of the model, we note here that during preoprocessing we normalize all input IFU maps individually, essentially removing any global information present. Thus, the input images only contain the gradients of the maps and are stripped of all mass dependencies. The model is then trained to learn relations between the gradients and the global ex-situ stellar mass fractions. \looseness-2

\subsection*{MaNGA}
MaNGA\cite{2015ApJ...798....7B} (Mapping Nearby Galaxies at Apache Point Observatory) is the integral field spectroscopic survey of the Sloan Digital Sky Survey IV (SDSS-IV), designed to systematically map the kinematics and spatially resolved spectroscopic properties of nearby galaxies. MaNGA provides spatially-resolved spectra for over 10,000 galaxies with stellar masses in the range of $10^{9} M_\odot < M_\star < 10^{12} M_\odot$. The survey covers redshifts in the range $0.01 < z < 0.15$ (with a median $z \sim 0.03$) and a wavelength range 3600–10300 Å at a resolution of R $\sim$ 2000. 
Every galaxy is observed using fiber bundles with diameters that vary from 12'' (19 fibers) to 32'' (127 fibers). The complete MaNGA sample consists of three sub-samples: the Primary sample, offering optimized spatial coverage out to 1.5Re ($\sim 50\%$); the Secondary sample, extending to a radial coverage of 2.5Re ($\sim 33\%$); and the Color-Enhanced sub-sample, which supplements the Primary sample in poorly sampled regions of the NUV-i vs $M_i$ color-magnitude plane ($\sim 17\%$) \cite{2017AJ....154...86W}. MaNGA offers a statistically significant sample of galaxies across diverse stellar masses and environments. To move to a volume-complete sample, we use the volume correction weights provided by ref\cite{2017AJ....154...86W}. In the current work, the stellar mass measurements assumed are obtained from the NASA-Sloan Atlas (NSA) catalog, derived through K-correction fitting for elliptical Petrosian fluxes. We additionally apply a correction on the NSA stellar masses by the cosmological parameter $H_0 = h 100 \textrm{km/s Mpc}$ to match the simulations. The h constant assumed in the NSA catalogue is $h = 1$, while the one
used in the simulations corresponds to Planck Collaboration 2015\cite{2016A&A...594A..13P}, with a value of h = 0.6774. The correction applied (in logscale units of solar masses) is the following:
    $M^\star_{corr} = M^\star - 5\textrm{log}_{10}(h)/2.5$.

\subsection*{MaNGIA}
MaNGIA\cite{2023A&A...673A..23S} is a forward-modelled sample of galaxies from the TNG50 simulation \cite{2019MNRAS.490.3196P, 2019MNRAS.490.3234N}, the highest-resolution simulation of the IllustrisTNG suite. The simulated galaxies from TNG50 are chosen to match MaNGA galaxies in stellar mass, size and redshift and are forward-modelled so that a mock MaNGA-like datacube is constructed for each simulated galaxy. This is performed by an extended processing pipeline, which includes the generation of the mock spectra from the simulation particles, the replication of the fibre bundle configuration to mimic MaNGA, as well as the addition of observational effects, such as  atmospheric seeing and noise. Afterwards, the resolved mock datacubes are analyzed with the PIPE3D pipeline \cite{2022ApJS..262...36S, 2022NewA...9701895L}, offering mock stellar population and kinematic maps. The PIPE3D analysis code is one of the procedures used on the official MaNGA release, allowing for a direct comparison of the mock data sample with observations. MaNGIA successfully reproduces overall trends in metallicity, age, and kinematics observed in MaNGA, rendering it an  ideal sample for machine learning applications, as the one presented in the current work. We note that, in addition to the publicly available MaNGIA galaxy sample, we have augmented our training set with an additional sample of $\sim$ 1600 MaNGIA galaxies, which are not yet included in the public dataset. \looseness-2

\subsection*{Neural Network model architecture}
We wish to infer the ex-situ stellar mass fraction of galaxies from 2D spatially-resolved maps of stellar mass density, stellar velocity and velocity dispersion. Since our training dataset (MaNGIA) derives from a simulation, where the ground truth is available, a supervised machine learning approach is considered suitable for this task. However, MaNGIA is not created with the current application under consideration and as a result the original dataset is very unbalanced in terms of the distribution of the ex-situ stellar mass fractions. More specifically, the vast majority of the mock galaxies in MaNGIA have a very low ex-situ stellar mass fraction and only a handful of galaxies have a fraction of accreted mass higher than 40\%. This can cause a traditional supervised neural network to predict only biased low values. To circumvent this issue, we turn to a self-supervised architecture. \looseness-2

Self-supervised learning is an approach of machine learning, lately becoming increasingly popular in astrophysical applications, that allows the extraction of meaningful representations from unlabeled data by solving pretext tasks. These representations can then be used for a variety of downstream tasks. In our case, we select a self-supervised learning approach, as it has been shown to help with unbalanced data, by providing representations that are more robust to dataset imbalance than supervised learning \cite{2021arXiv211005025L}. We decide to use the Bootstrap your own latent (BYOL) self-supervised learning approach \cite{2020arXiv200607733G}, which has been shown to extract robust representations even from limited samples. The BYOL model uses two neural networks with the same architecture that receive different augmented views of the same input image and output a representation in the embedding latent space. The network is trained to minimize the distance between the two embeddings ($z_i$ and $z_i^+$), effectively learning robust representations of the images that are invariant to the augmentations applied. To further assist our model in capturing relevant features in the data that correlate with the ex-situ stellar mass fraction, we additionally add a regressive head ($q_r$) to the self-supervised network during training, designed to learn the accreted stellar mass fractions from the representations. As a result, the loss function of the model is a combination of the BYOL loss term and the regression loss (mean-squared error):

% \cite{2023RASTI...2..441H} review of self-supervised

\begin{equation}
    \centering
    \mathcal{L} = \mathcal{L}_{BYOL} + \lambda\mathcal{L}_{regr} = 2 - 2\frac{ \langle z_i, z_i^+ \rangle}{\lVert z_i\rVert_{2} \cdot \lVert z_i^+\rVert_{2} } + \lambda \frac{1}{n} \sum (Y_i - q_r(z_i) )^2
\end{equation}

To further enhance the performance of the model, we artificially balance our training dataset prior to training by including simple augmentations in the data, specifically targeted for galaxies with high ex-situ stellar mass fractions. The said augmentations include random cropping, resizing, horizontal and vertical flipping. The model architecture is implemented in Keras with a Tensorflow backend component. We train our models on a NVIDIA Tesla P100 GPU for an optimized execution time.

\subsection*{Uncertainties of the model}

\begin{figure}
    \centering
    \includegraphics[width=0.43\linewidth]{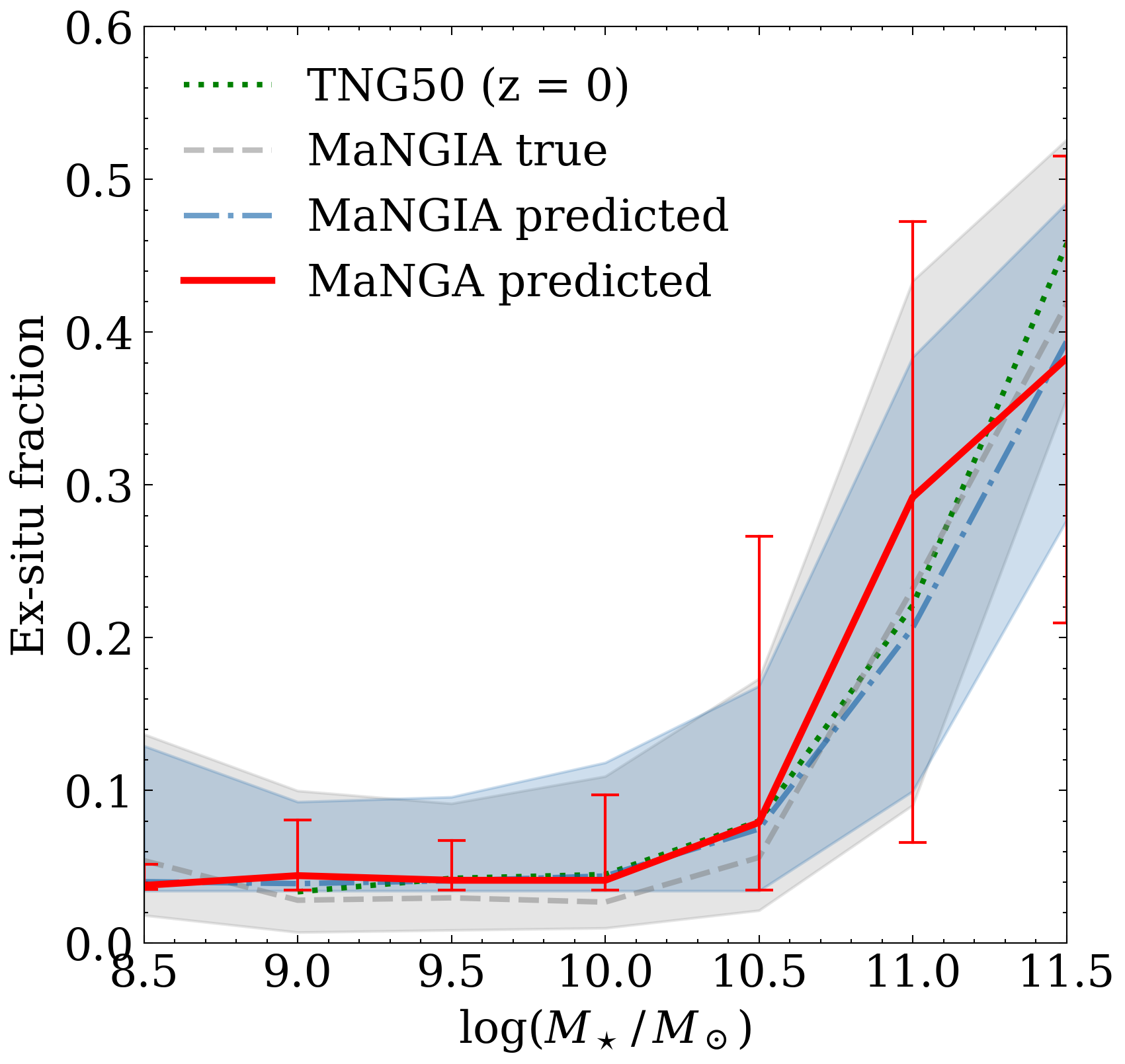} 
    \hspace{0.5cm}
    \includegraphics[width=0.41\linewidth]{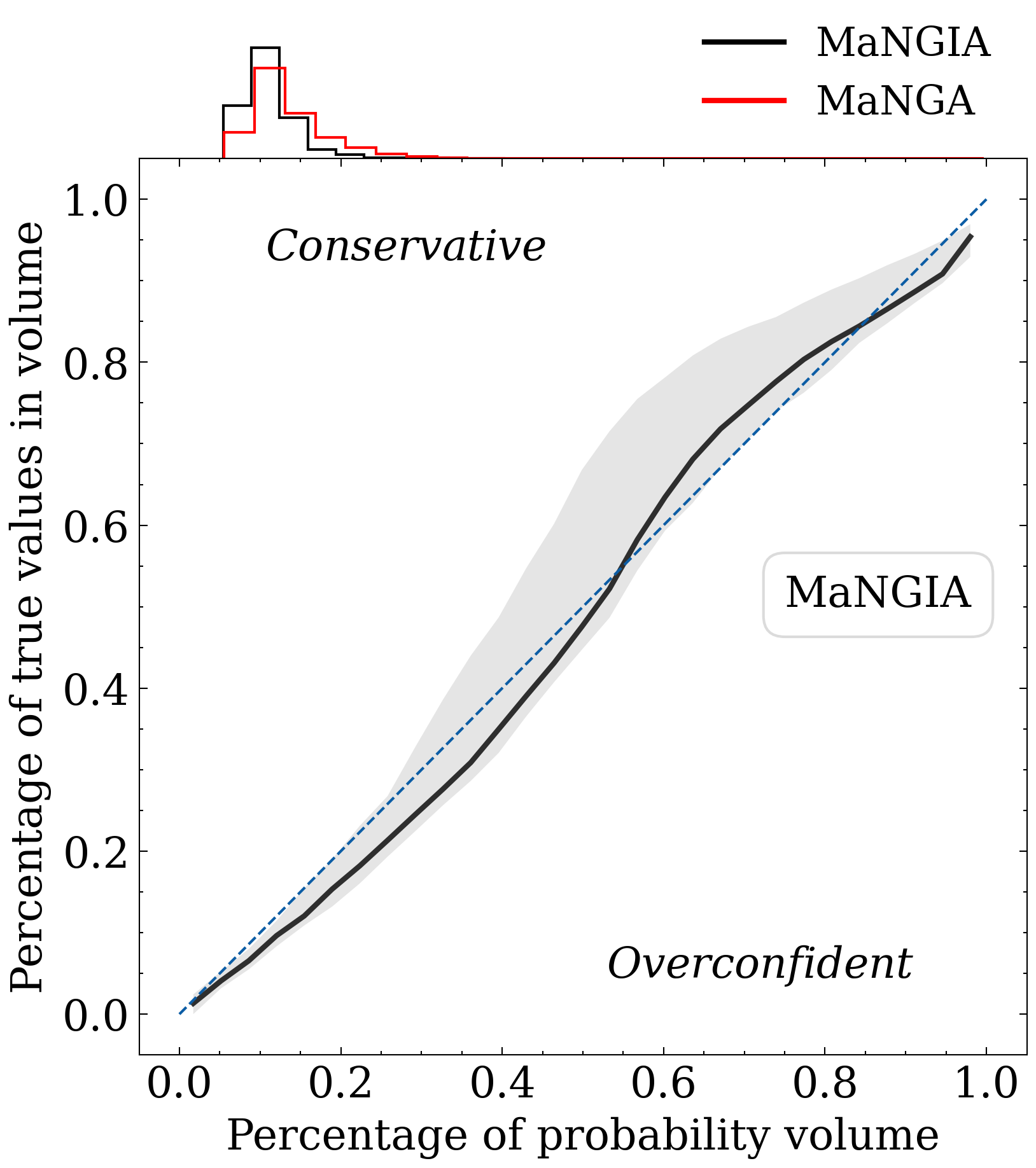} \hfil        

    \caption{Validation of the NN model and its uncertainties on the MaNGIA mock dataset and the MaNGA real galaxies. (a) The recovery of the median ex-situ stellar mass fraction vs. stellar mass relation for $\sim$ 10,000 MaNGA galaxies (red solid line) as it is predicted from the ensemble of the NN models trained on the MaNGIA mock dataset. For a visual validation of the model, both the ground-truth (grey dashed line) and the model predictions (blue dash-dotted line) are shown for the ex-situ fractions of the MaNGIA test set, revealing a close recovery of the underlying relation. The median relation for the TNG50 simulation, from which the MaNGIA sample originates, at redshift $z = 0$ is shown with a green dotted line. The shaded regions and error bars contain the 68 per cent of all data. (b) The coverage statistical test applied on the MaNGIA test set. The black solid line corresponds to the median of the uncertainty calibration of the models trained on MaNGIA. The perfect calibration for the uncertainties is shown with a diagonal dashed blue line. Our models seem to follow closely the perfect calibration, which is the desired behavior. On the top, we also show the distribution of the produced ensembled uncertainties of the models on MaNGIA and MaNGA and find that they are comparable, with the model producing slightly higher uncertainties on the unknown MaNGA dataset.}
    \label{fig:uncertainties}
\end{figure}

Prior to applying the trained model to the MaNGA dataset, we execute a set of validation steps to ensure the network's recovery on the MaNGIA mock dataset as well as to calibrate the model's uncertainties. First, we adopt an ensemble approach in an attempt to reduce epistemic uncertainties and enhance the reliability of the predictions of the model on MaNGA. More specifically, instead of depending on a single trained model, we train five neural networks, each with the architecture detailed in the previous section but using distinct weight initializations. Employing this ensemble strategy enables us to combine the predictions from the five models for each galaxy, effectively reducing uncertainty and yielding more robust predictions. We find that this approach indeed results to an increased accuracy on the unseen test dataset from MaNGIA, where the ground-truth is available and accuracy is thus measurable. In the left panel of Fig. \ref{fig:uncertainties}, the recovery of the ex-situ stellar mass fractions is shown for the MaNGIA test set, along with the ensembled predictions of the models for the whole MaNGA sample. The predictions for MaNGIA trace closely the ground truth, enhancing our confidence on the results recovered on MaNGA.

Moreover, an essential consideration for evaluating the robustness of a machine learning model when applied to real galaxies involves providing a measure of uncertainty for the model predictions. In our approach, we address this crucial aspect through a second step of training: fine-tuning the neural network to predict a probability distribution instead of a single value for the ex-situ stellar mass fraction of each galaxy. This involves freezing the representations learned from the self-supervised training and exclusively training the model to effectively map these representations to a posterior distribution of the ex-situ stellar mass fraction. By sampling from the posteriors of the five fine-tuned models for each galaxy, we can ensemble the predicted distributions from all samples and utilize the produced standard deviation as a confidence measurement of the models on each prediction. \looseness-2

To establish the reliability of the standard deviation as a measure of uncertainty, we conduct a statistical coverage test on the models' predictions for the MaNGIA test set. This test assesses the accuracy of predicted posteriors by determining the percentage of times the error falls within a certain percentage of the posterior width. Ideally, this percentage should align with the specified width (e.g., 68 percent of galaxies should have a prediction error $\leq 1 \sigma$ for a 68 percent probability volume). Our findings indicate that our models exhibit close-to-perfect calibration while being slightly conservative, as it can be seen in the right panel of Fig. \ref{fig:uncertainties}, underlining that they indeed provide meaningful measures of uncertainty. As the ground-truth is not available for MaNGA galaxies, the same coverage test cannot be applied. However, we find that the uncertainty measurements the models provide for MaNGIA and MaNGA (top of the right panel of Fig. \ref{fig:uncertainties}) are comparable, with MaNGA exhibiting slightly higher uncertainties, as it was expected. We note here that by fine-tuning the models, we end up with a handful of galaxies in MaNGA that have predicted distributions that lie outside the prior space for ex-situ stellar mass fractions (negative mean values). However, by removing them and re-applying the statistical analysis of the current work, we find that the same trends arise for galaxy characteristics and environment. Thus, these discrepancies do not highly affect the produced results, and might even indicate possible outliers in the observations that are not sufficiently covered in the simulated space.

\subsection*{Implications from the ex-situ definition}
An important consideration that needs to be addressed in our work is the definition of in-situ versus ex-situ stellar mass in the training set. As MaNGIA originates from the IllustrisTNG (TNG50) cosmological simulation, the ground-truth provided during training for the ex-situ stellar mass is carried over from that simulation. More specifically, the definition of the ex-situ stellar mass fraction is derived from ref.\cite{2016MNRAS.458.2371R}, where in-situ stars are considered all particles formed in galaxies that lie along the main progenitor branch of the galaxy in which they currently reside. All stellar particles formed outside the main progenitor branch are considered ex-situ stars. The said definition results in the primary growth of ex-situ stellar mass predominantly occurring at $z=2$ and beyond in the Illustris cosmological model \cite{2016MNRAS.458.2371R}. However, this may be at odds with observations suggesting that massive early-type galaxies experience most of their stellar mass and size growth at lower redshifts ($z < 2$) \cite{2005ApJ...626..680D, 2008ApJ...677L...5V, 2014ApJ...788...28V}. Nevertheless, in the IllustrisTNG simulations, specifically for massive galaxies, the relative contribution of the ex-situ fraction to the total stellar mass also exhibits an increasing trend at $1 < z < 2$ \cite{2016MNRAS.458.2371R}. This trend can rise the ex-situ stellar mass fraction from 20\% to 50\% for galaxies with a stellar mass of $M_\star \sim 10^{11} M_\odot$ \cite{2019MNRAS.487.5416T}, thereby more consistent with observational results within this stellar mass regime.

\subsection*{Robustness of the predictions}

While simulation-based inference is key for predicting unobservable properties of galaxies, it can be affected by systematic uncertainties from using a specific simulated dataset for training, raising concerns about the model's generalization capabilities. In our case, assessing the model's robustness requires examining multiple levels, each with its potential sources of bias: the raw simulation space, the mock space, and the observational space.

\begin{figure}
    \centering
    \includegraphics[width=0.33\linewidth]{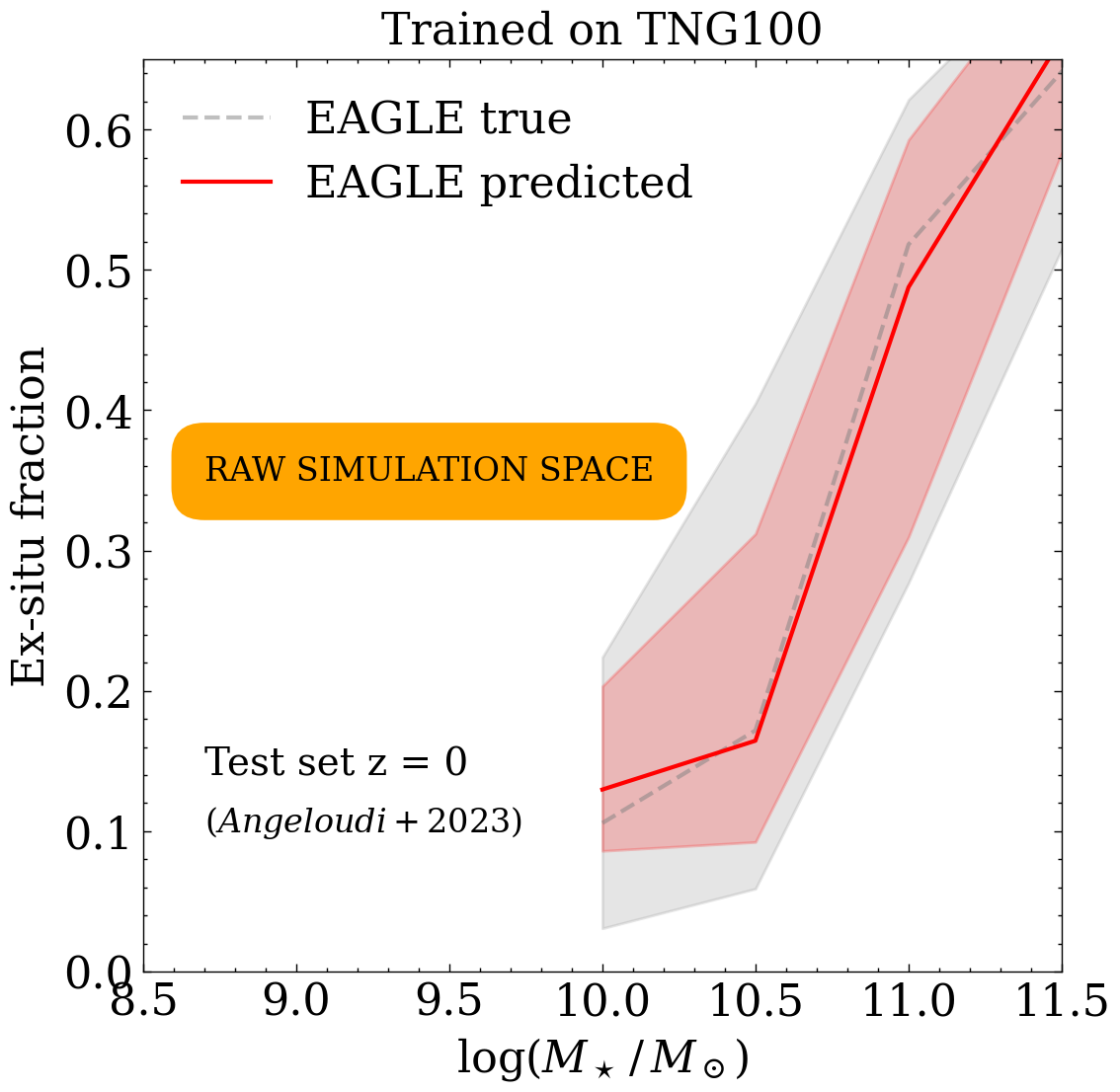} 
    %\hspace{0.35cm}
    \includegraphics[width=0.33\linewidth]{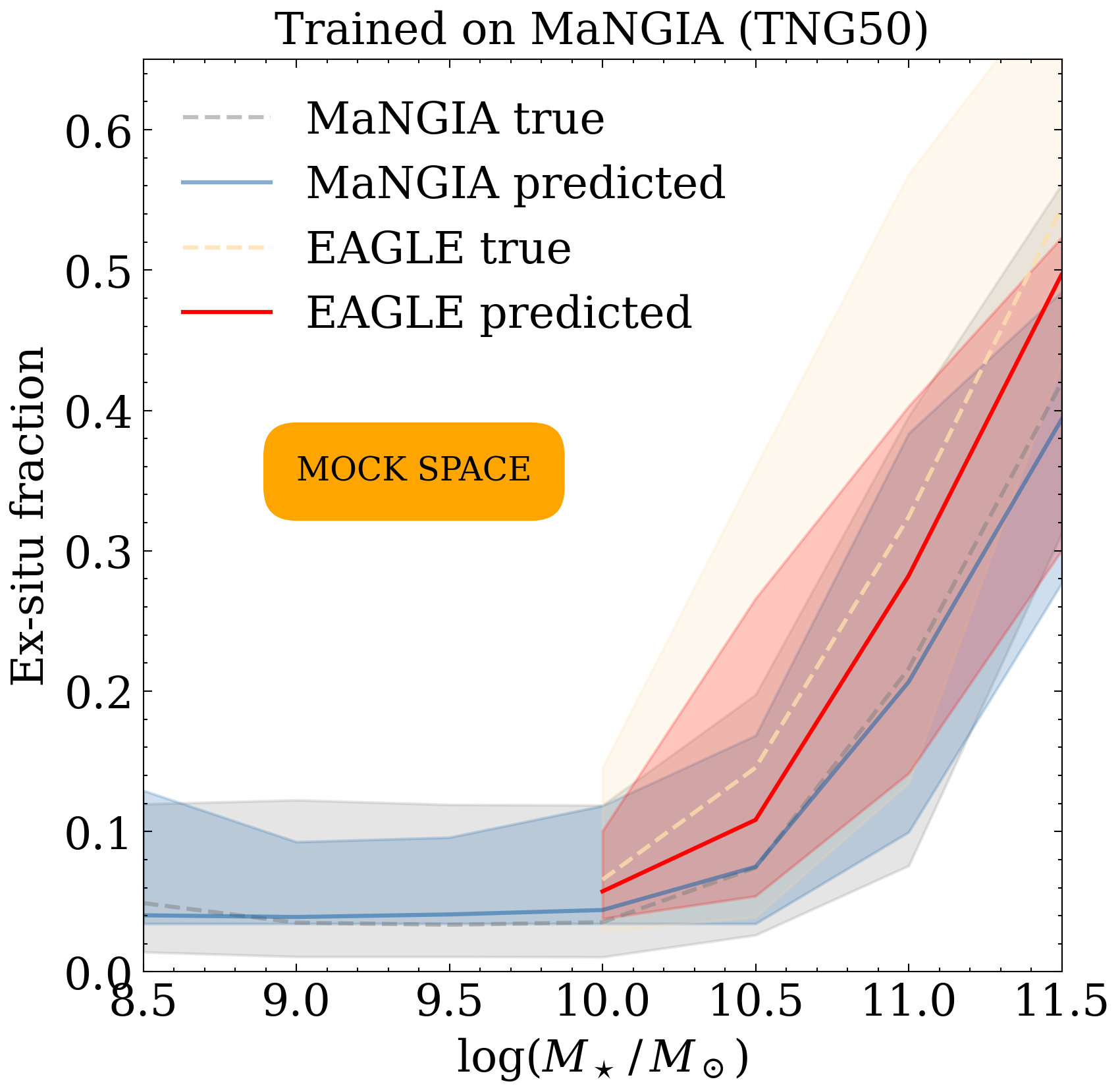}
    %\\\vspace{0.5cm}
    %\includegraphics[width=0.44\linewidth]{Figures/ExsituF_vs_total_mass_mangia_manga_eagle_mock_effect.png} 
    \includegraphics[width=0.33\linewidth]{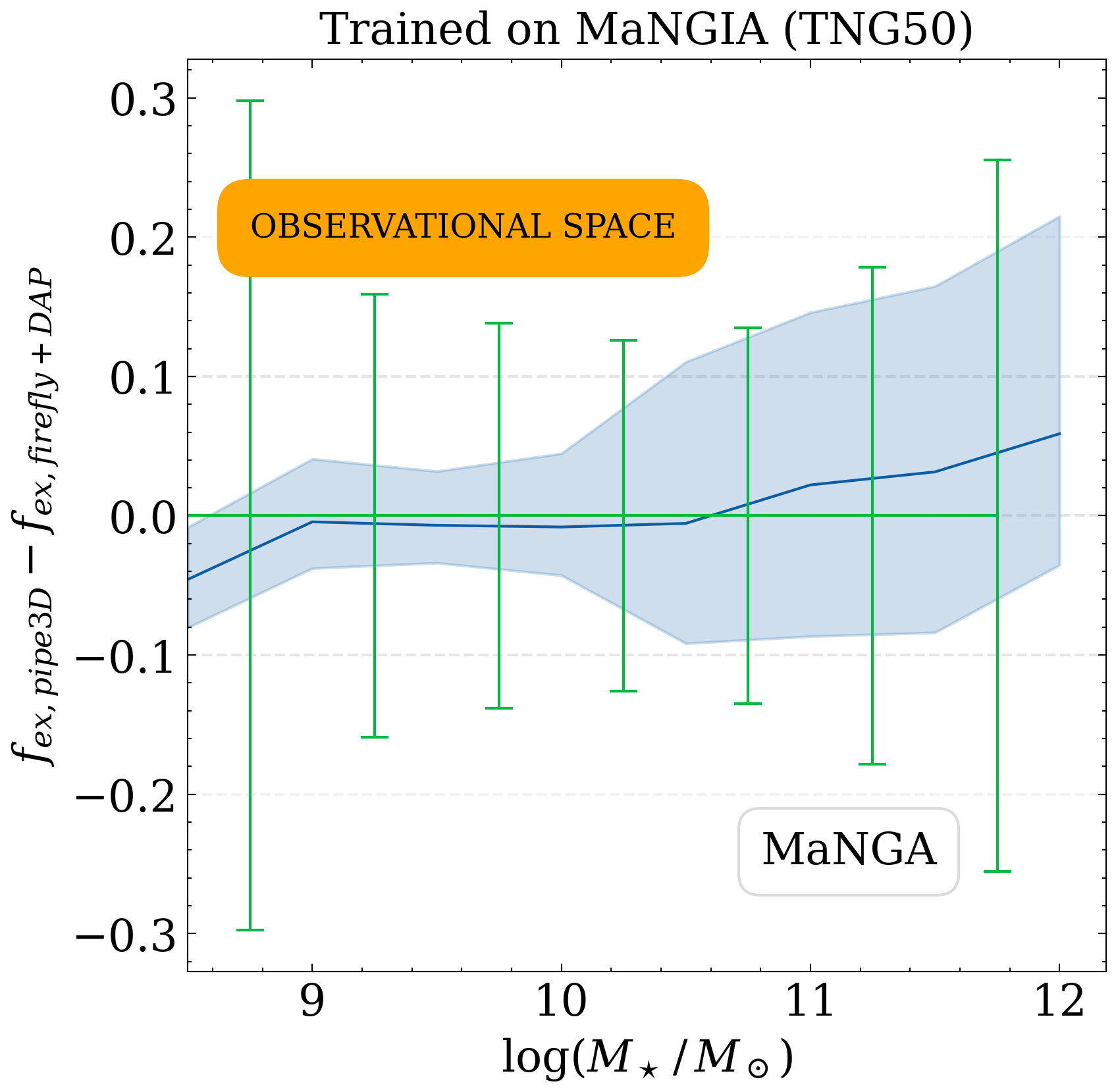}   

    \caption{Measurements of robustness for the NN model's predictions in the raw simulation space, mock space, and observational space. (a) In the raw simulation space, we demonstrate the model's robustness across different sub-grid physics, as shown in \cite{2023MNRAS.523.5408A}. A model trained on TNG100 successfully recovers the ex-situ vs. stellar mass relation on a test set from the EAGLE simulation at $z=0$, using IFU-like maps of stellar mass and kinematics, equivalent to the current work. (b) In the mock space, a model trained on MaNGIA (TNG50) can recover on a satisfactory level the ex-situ vs. stellar mass relation for approximately 150 EAGLE galaxies at $z=0.1$, mocked with the same pipeline as the MaNGIA training set. (c) In the observational space (MaNGA), we compare predictions from two different products/pipelines (PIPE3D vs. Firefly + MaNGA DAP) for acquiring the IFU input maps. The median difference is plotted with a solid blue line, and the shaded region contains 68\% of the data. Additionally, the mean standard deviation of the models when applied on MaNGA PIPE3D is shown with the green errorbars in each stellar mass bin. The median difference between the predictions for the different MaNGA pipelines is nearly zero across the stellar mass range and falls within the uncertainties produced by the models.}
    \label{fig:robustness}
\end{figure}

Focusing initially on the raw simulation space, the primary source of bias originates from the particular sub-grid physics of the training simulation, as the model may learn patterns tied to the underlying galaxy-formation model, which do not generalize to different sub-grid physics. To address this issue, we build upon our recent work \cite{2023MNRAS.523.5408A}, which identified a set of inputs robustly predicting the ex-situ stellar mass fraction across two cosmological simulations. Specifically, we found that IFU-like maps of stellar mass, stellar velocity, and velocity dispersion can accurately predict the ex-situ stellar mass fraction across the TNG100 and the EAGLE cosmological simulations, while other stellar population properties like age and metallicity induce bias. Demonstrating this robustness, Fig. \ref{fig:robustness} (a) illustrates the recovery of the ex-situ vs. stellar mass relation for a test set from the EAGLE cosmological simulation at $z=0$ predicted by a NN trained on TNG100. This showcases that the NN model generalizes well to different sub-grid physics and therefore is not anchored to a specific model. Thus, as an initial step towards generalization, we strategically limit the inputs supplied to the neural network in the current work to this robust subset of IFU maps. \looseness-2

%(c) The effect of different mocking choices in the MaNGIA pipeline. Different methods for converting galaxy size from 3D ($R^{3D}_\star$) to 2D (half-light radius) affect the signal-to-noise ratio and, consequently, the inferred results for the EAGLE mock dataset when using a model trained on MaNGIA. 

The process of mocking, i.e., translating simulation outputs to the observational space, also requires assumptions that can introduce systematic uncertainties. To properly quantify this effect, we should ideally perform a similar cross-training of the neural network model among different mock datasets and validate the inferred results on the MaNGA sample. However, this requires at least two different mock samples, each with realism matched to MaNGA and originating from distinct cosmological simulations. Unfortunately, apart from MaNGIA (which originates from TNG50), there is no comparable mock dataset from a different cosmological simulation and creating one from scratch is a very computationally expensive task, beyond the scope of the current work. Nonetheless, since the MaNGIA mocking pipeline is publicly available, we decided to create a limited mock dataset from the EAGLE cosmological simulation solely for testing purposes. We produced a limited sample of approximately $\sim150$ mock galaxies at $z = 0.1$, spanning a stellar mass range of $10^{10} M_\odot < M_{\star} < 10^{11.5} M_\odot$. We then applied the MaNGIA-trained model on the EAGLE mock sample and plot the results in Fig. \ref{fig:robustness} (b). We find that the model satisfactorily reconstructs the stellar mass vs. ex-situ fraction relation, despite being trained on a higher resolution cosmological box and different sub-grid physics. Although there is an offset of about 5\% between the true and predicted relations, the model is capable of closely recovering the slope of the true relation, even though it is distinctly different from the one it was trained on. This is a reassuring result, as it indicates that the model is not merely projecting trends from the TNG50 simulation, but is indeed learning generalized mappings between the radial gradients and the ex-situ stellar mass fractions. 

We note here that the offset between the true and the predicted relations for the EAGLE sample is only observed in the mock space and not the raw simulation space. This can be partially attributed to the different resolutions between MaNGIA (TNG50) and EAGLE/TNG100, as well as the mocking procedure itself. Specifically, calculating a projected measurement of galaxy size in the 3D simulation space equivalent to the half-light radius used for the fiber recombination in MaNGA is challenging and can affect the noise in the resulting mock IFU maps. For the MaNGIA mock dataset, the authors used the radius of a sphere containing half of its stellar mass ($R^{3D}\star$) and fitted it to the observed radius from synthetic images of TNG50 galaxies \cite{2019MNRAS.483.4140R}, using the semi-major axis of the ellipse containing half of the stellar light ($R^{Petro}_{0.5}$). We used the same fit for the EAGLE mock sample, but this conversion may not be optimal for EAGLE galaxies, affecting the level of noise in the produced maps and thus contributing to the offset observed in the mock space.

Finally, reaching the observational space, we aim to examine the impact of the processing pipeline used for the recovery of the MaNGA IFU maps on the predictions of the model. For that, we compare the inferred ex-situ stellar mass fractions recovered from the PIPE3D maps with those recovered using IFU maps from a different pipeline. Specifically, we use the products from the Firefly analysis code \cite{2017MNRAS.472.4297W} to acquire stellar mass density maps and the MaNGA data-analysis pipeline (MaNGA DAP) \cite{2019AJ....158..231W} for the stellar kinematic maps, as an alternative to the PIPE3D maps. In Fig. \ref{fig:robustness} (c), we plot with a blue solid line the median of the difference between the predictions acquired for each galaxy in MaNGA when the distinct pipelines are used to resolve the IFU input maps across the stellar mass range. We find that the models are capable of robustly predicting the ex-situ stellar mass fraction of a galaxy independently of the processing pipeline that was used to derive the stellar maps. There exists an increase in the difference of the predictions for the most massive galaxies, but it is accompanied with an higher uncertainty produced by the models (green error bars). This further strengthens our confidence in the predicted results. As a final validation test, we also attempted to train the neural network on a different mock MaNGA dataset (iMaNGA \cite{2023MNRAS.522.5479N}) for a direct comparison of the inferred results, but the sample's limited size ($\sim$ 1000 galaxies) did not allow for an effective training of the model.

\subsection*{MaNGA resolved properties and environment}

To unveil the secondary dependecies of the ex-situ stellar mass, we use properties that have already been measured for the MaNGA survey in previous works. For the morphological type that each galaxy belongs to, we use the MaNGA PyMorph morphological catalogue \cite{2022MNRAS.509.4024D}, which provides deep-learning-based morphological classifications for the SDSS Data Release 17. We separate the MaNGA galaxies into 2550 ellipticals, 966 S0-type and 5428 spiral galaxies by using the following recommended restrictive selection:

\begin{itemize}
    \item Spiral galaxies: T-Type $ > 0$ AND $P_{LTG} \geq 0.5$ AND VC = 3
    \item S0 galaxies: T-Type $ \leq 0$ AND $P_{S0} > 0.5$ AND $P_{LTG} < 0.5$ AND VC = 2
    \item Early-type galaxies: T-Type $\leq 0$ AND $P_{S0} \leq 0.5$ AND $P_{LTG} < 0.5$  AND VC = 1
\end{itemize}
where VC corresponds to a visual classification (VC=1 for ellipticals, VC=2 for S0, VC=3 for LTG and irregulars, and VC=0 for unclassifiable galaxies) and $P_{class}$ corresponds to the probability a galaxy belongs to that class as outputted from the machine-learning model. 

\begin{figure}
    \centering
    \includegraphics[width=0.48\linewidth]{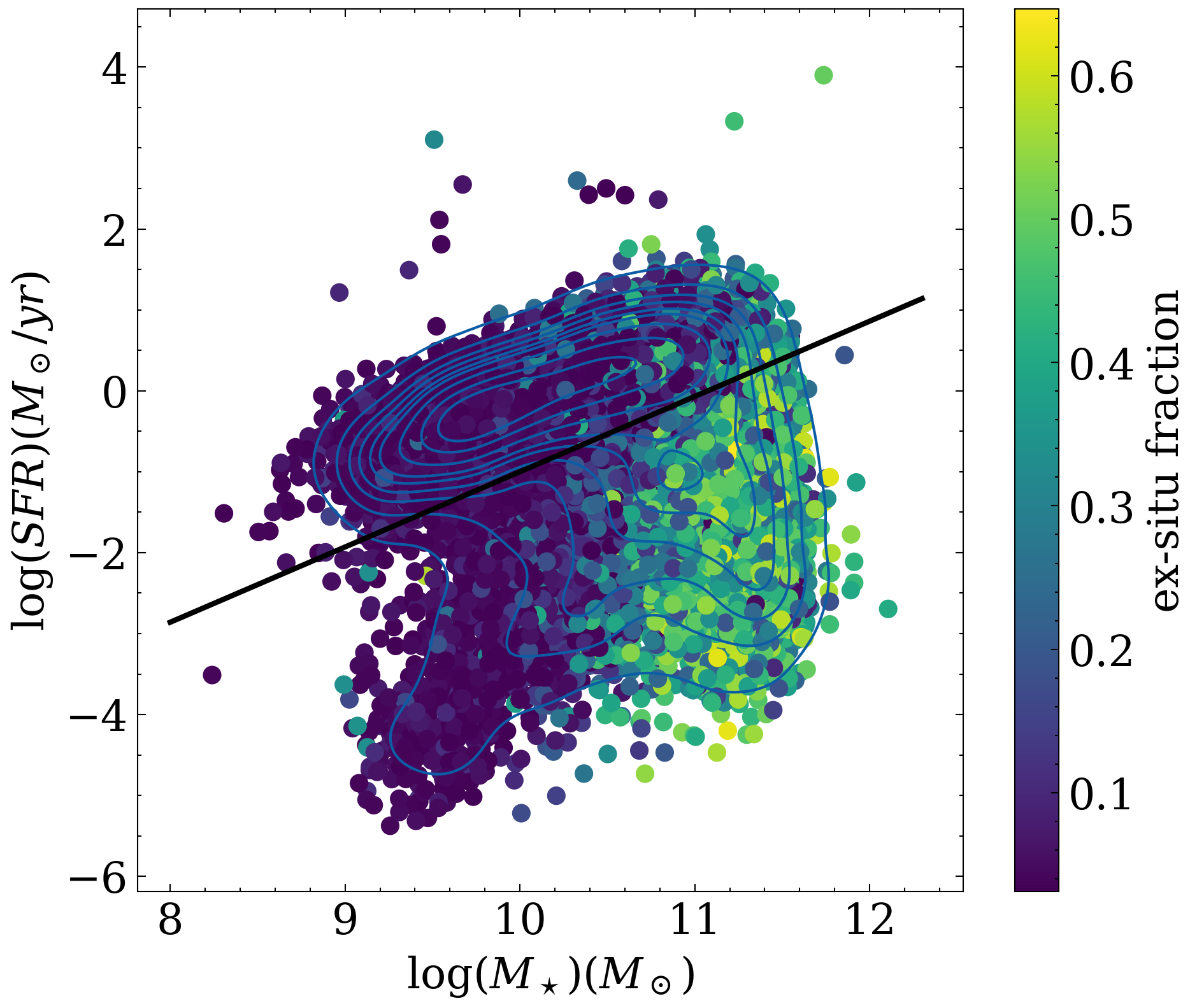
    } \hfil
    \includegraphics[width=0.48\linewidth]{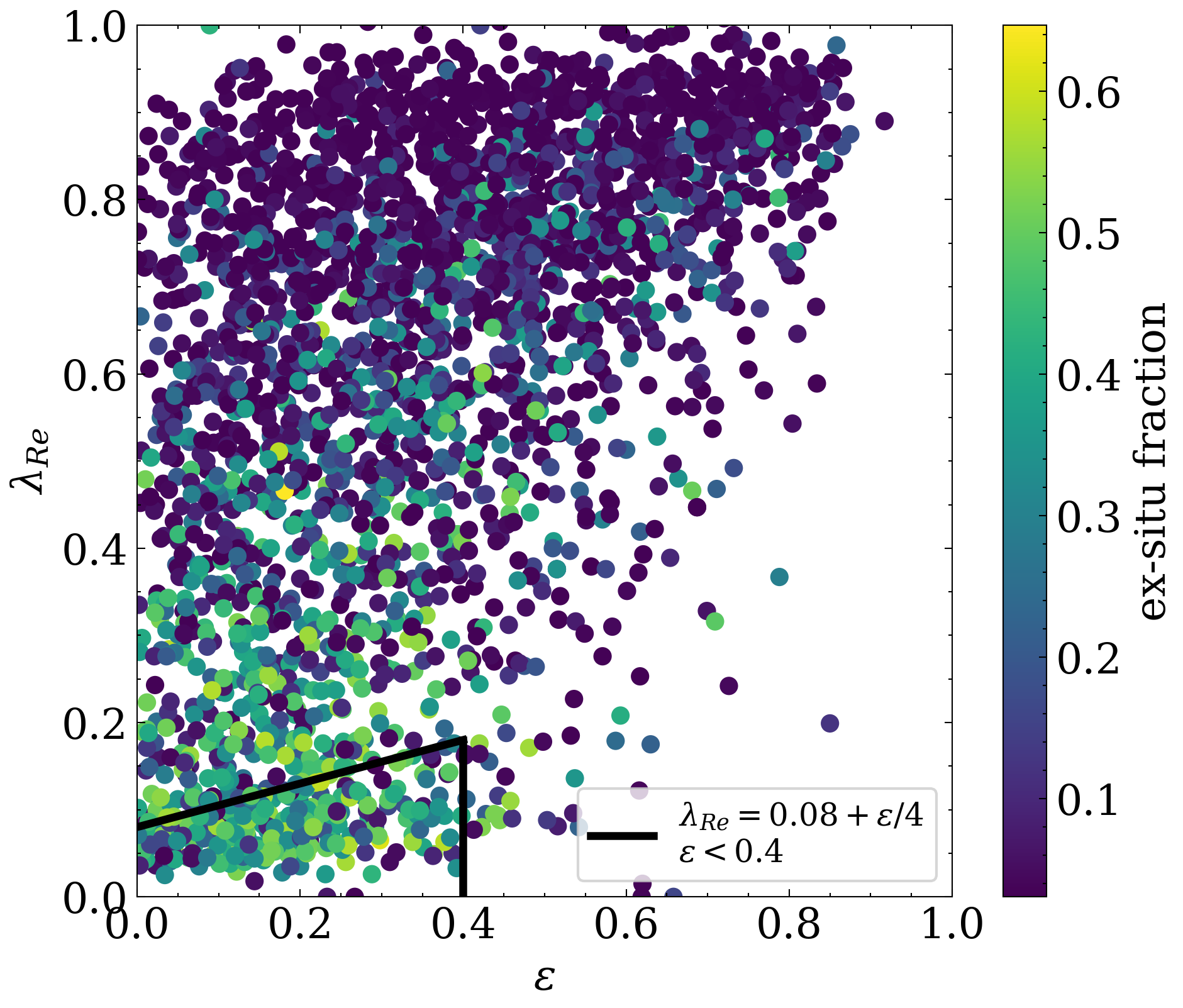}
    
    \caption{The two planes used for separating the MaNGA sample into different subgroups based on star-formation activity and rotation. (a) The classification for quenched and star-forming galaxies using the SFR from Ha from PIPE3D on MaNGA galaxies. The black solid line separates the star-forming from the quiescent galaxies. The points are color-coded with the resolved ex-situ stellar mass fraction, revealing an higher abundance of higher accreted fractions in the quenched population. (b) The classification for slow and fast rotators for a limited sample of 2300 galaxies from MaNGA using the criterion provided in the legend for slow rotators. The points are color-coded with the resolved ex-situ stellar mass fraction, revealing a dependency of the accreted mass on angular momentum.}
    \label{fig:sfr}
\end{figure}

To classify the MaNGA galaxies regarding their star-formation activity into quenched and star-forming, we use the PIPE3D analysis \cite{2016RMxAA..52..171S}
applied to the DR17 data release of MaNGA. We obtain the integrated star-formation rate (SFR) derived from the Ha emission line from the said analysis and we separate our galaxies into star-forming and quiescent on the SFR vs. stellar mass plane. In Figure \ref{fig:sfr}, the classification cut used in the current work is shown with a black solid line. The points are additionally color-coded with the resolved ex-situ stellar mass fraction, revealing already from this plot a secondary dependency of the accreted stellar mass to the star-formation activity of the galaxy. This separation criterion produces a sample of 5696 star-forming and 4519 quenched galaxies.  

In terms of rotation, we use the angular momentum proxy $\lambda_{Re}$ and the ellipticity $\epsilon$ measurements for about 2300 MaNGA galaxies provided by \cite{2018MNRAS.477.4711G}. Plotting these galaxies in the angular momentum vs. ellipticity plane, we employ the classification criterion proposed by \cite{2016ARA&A..54..597C} to distinguish slow and fast rotators, depicted in the right panel of Fig. \ref{fig:sfr}. Following this criterion -- labeling galaxies with $\lambda_{Re} < 0.08 + \epsilon$ and $\epsilon < 0.4$ as slow rotators and the remainder as fast rotators -- we identify a subset of 268 slow rotators and 2520 fast rotators. Notably, this plane also exhibits a discernible trend of ex-situ stellar mass with angular momentum, as indicated by the color-coded representation in the figure.

Finally, for a measurement of the host halo environment properties, we turn to the self-calibrating halo-based group finder applied on SDSS by \cite{2021ApJ...923..154T}. This catalog provides, among others, the likelihood ($P_{sat}$) of each galaxy being a satellite, the dark matter halo mass ($M_{halo}$), and the count of satellites within the host halo ($N_{sat}$) for central galaxies. We use a relatively strict criterion to separate our MaNGA galaxies into centrals ($P_{sat}$ < 0.1) and satellites ($P_{sat} \geq $  0.9). We additionally remove isolated galaxies from our sample (central galaxies that have no satellites) to ensure more reliable halo mass estimates. We end up with 2234 central galaxies and 1508 satellite galaxies. The range of the halo masses provided by the group catalog for the sample of central galaxies is $10^{11} M_\odot < M_{halo} < 10^{15} M_\odot$, peaking at $M_{halo} \sim 10^{13} M_\odot$. The majority of central galaxies in this analysis reside in Milky Way sized haloes and groups, while a small subsample lives in clusters ($\sim 150$ galaxies with $M_{halo} > 10^{14} M_\odot$).

\section*{Code availability}
\noindent Data directly related to this publication and its figures will be made available on request from the corresponding author. The MaNGA products (https://www.sdss4.org/dr17/manga/) and the MaNGIA dataset (https://www.tng-project.org/data/docs/specifications) are all publicly available. Upon acceptance, we plan to release the associated code, datasets and ML models developed here on GitHub.

\bibliography{sample}
\bibliographystyle{naturemag}

\section*{Corresponding Author} 
\noindent
Correspondence to Eirini Angeloudi (eirini@iac.es).

\section*{Acknowledgements} 
\noindent
EA thanks C. Gallart, S. Faber, K. Bundy and L. Scholz-D\'iaz for their insightful discussions. JFB and EA acknowledge support the Spanish Ministry of Science, Innovation and Universities (Grant Nos. PID2019-107427GB-C32 and PID2022-140869NB-I00) and through the project TRACES from the Instituto de Astrofísica de Canarias,  which is partially supported through the state budget and the regional budget of the Consejer\'ia de Econom\'ia, Industria, Comercio y Conocimiento of the Canary Islands Autonomous Community. MHC, RS and EA acknowledge financial support from the State Research Agency (AEI\-MCINN) of the Spanish Ministry of Science and Innovation under the grants ``Galaxy Evolution with Artificial Intelligence" with reference PGC2018-100852-A-I00 and "BASALT" with reference PID2021-126838NB-I00. This research made use of computing time available on the high-performance computing systems at the Instituto de Astrofisica de Canarias. The authors thankfully acknowledge the technical expertise and assistance provided by the Spanish Supercomputing Network (Red Espanola de Supercomputacion), as well as the computer resources used: the Deimos-Diva Supercomputer and the LaPalma Supercomputer, located at the Instituto de Astrofisica de Canarias.

\section*{Ethics and inclusion}
\noindent
Our research project was conducted with integrity, respect, and inclusivity towards all members of the scientific community. We strive to create a safe, inclusive and supportive environment where everyone can contribute with their perspectives and ideas, regardless of their background or identity.

\section*{Author contributions statement}
\noindent
E.A. performed the data analysis and wrote the paper. E.A., M.H.-C. and J.F.-B. conceived and designed the project. A.B. provided substantial comments on the overall design of the project and the interpretation of the results. R.S. significantly assisted with the creation of the mock dataset from EAGLE cosmological simulation. L.E. and A.P. provided comments on the paper and helped improve its overall focus. All authors interpreted and discussed the results and commented on the paper.

% Must include all authors, identified by initials, for example:
% A.A. conceived the experiment(s), A.A. and B.A. conducted the experiment(s), C.A. and D.A. analysed the results. All authors reviewed the manuscript. 

\section*{Competing interests}
\noindent
The authors declare no competing interests.

% The corresponding author is responsible for providing a \href{https://www.nature.com/sdata/policies/editorial-and-publishing-policies#competing}{competing interests statement} on behalf of all authors of the paper. This statement must be included in the submitted article file.

\end{document}